\pdfoutput=1
\documentclass[11pt,twoside,a4paper,cmspaper,final,collab]{cms-tdr}
\def\svnVersion{c0500f5}\def\svnDate{2023/08/22}\def\cmsCernNoTag{CERN-EP-2022-177}\def\cmsCernDate{\today}\def\cmsMessage{Published in the Journal of High Energy Physics as \href{http://dx.doi.org/10.1007/JHEP07(2023)229}{\doi{10.1007/JHEP07(2023)229}.}}
\begin{document}\cmsNoteHeader{SMP-21-014}

\renewcommand{\cmslogo}{\includegraphics[height=2.33cm]{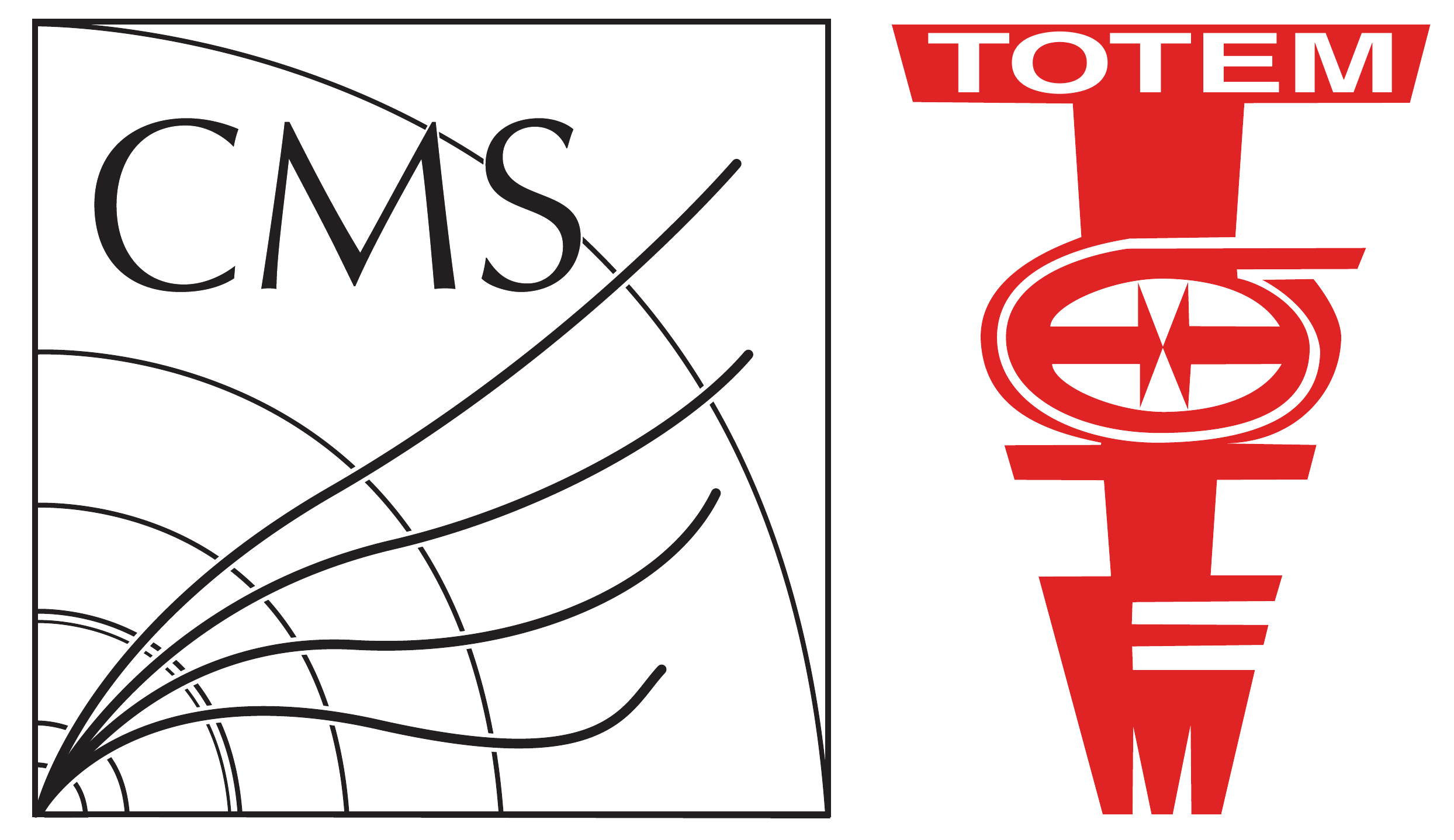}}
\renewcommand{\cmsCollabName}{The CMS and TOTEM Collaborations}
\renewcommand{\cmsTag}{CMS-\cmsNUMBER\\&TOTEM-2022-002\\}
\renewcommand{\cmsPubBlock}{\begin{tabular}[t]{@{}r@{}l}&CMS \cmsSTYLE\
\cmsNUMBER\\&TOTEM NOTE 2022-004\\\end{tabular}}
\newcommand{\VV}{\ensuremath{\mathrm{VV}}}
\newcommand{\Pj}{\ensuremath{\mathrm{j}}}
\newlength\cmsTabSkip\setlength{\cmsTabSkip}{1ex}
\providecommand{\cmsTable}[1]{\resizebox{\textwidth}{!}{#1}}
\renewcommand{\cmsCopyright}{\copyright\,\the\year\ CERN for the benefit of the CMS and TOTEM Collaborations.}

\cmsNoteHeader{SMP-21-014}
\title{Search for high-mass exclusive \texorpdfstring{$\PGg\PGg \to \PW\PW$}{gamma gamma to W W} and \texorpdfstring{$\PGg\PGg \to \PZ\PZ$}{gamma gamma to Z Z} production in proton-proton collisions at \texorpdfstring{$\sqrt{s}=13\TeV$}{sqrt(s) = 13 TeV}}

\author*[cern]{CMS and Totem Collaborations}

\date{\today}

\abstract{A search is performed for exclusive high-mass $\PGg\PGg \to \PW\PW$ and $\PGg\PGg \to \PZ\PZ$ production in proton-proton collisions using intact forward protons reconstructed in near-beam detectors, with both weak bosons decaying into boosted and merged jets. The analysis is based on a sample of proton-proton collisions collected by the CMS and TOTEM experiments at $\sqrt{s}=13\TeV$, corresponding to an integrated luminosity of 100\fbinv. No excess above the standard model background prediction is observed, and upper limits are set on the  
$\Pp\Pp\to\Pp\PW\PW\Pp$ and $\Pp\Pp\to\Pp\PZ\PZ\Pp$ cross sections in a fiducial region defined by the diboson invariant mass $m(\PV\PV)>1\TeV$ (with $\PV =\PW,\PZ$) and proton fractional momentum loss $0.04 < \xi < 0.20$. The results are interpreted as new limits on dimension-6 and dimension-8 anomalous quartic gauge couplings.}

\hypersetup{%
pdfauthor={CMS Collaboration},%
pdftitle={Search for high-mass exclusive gamma gamma to WW and gamma gamma to ZZ production in proton-proton collisions at sqrt(s) = 13 TeV},%
pdfsubject={CMS},%
pdfkeywords={CMS, PPS}}

\maketitle 

\section{Introduction}
\begin{figure}[ht]
\centering
\includegraphics[width=0.4\textwidth]{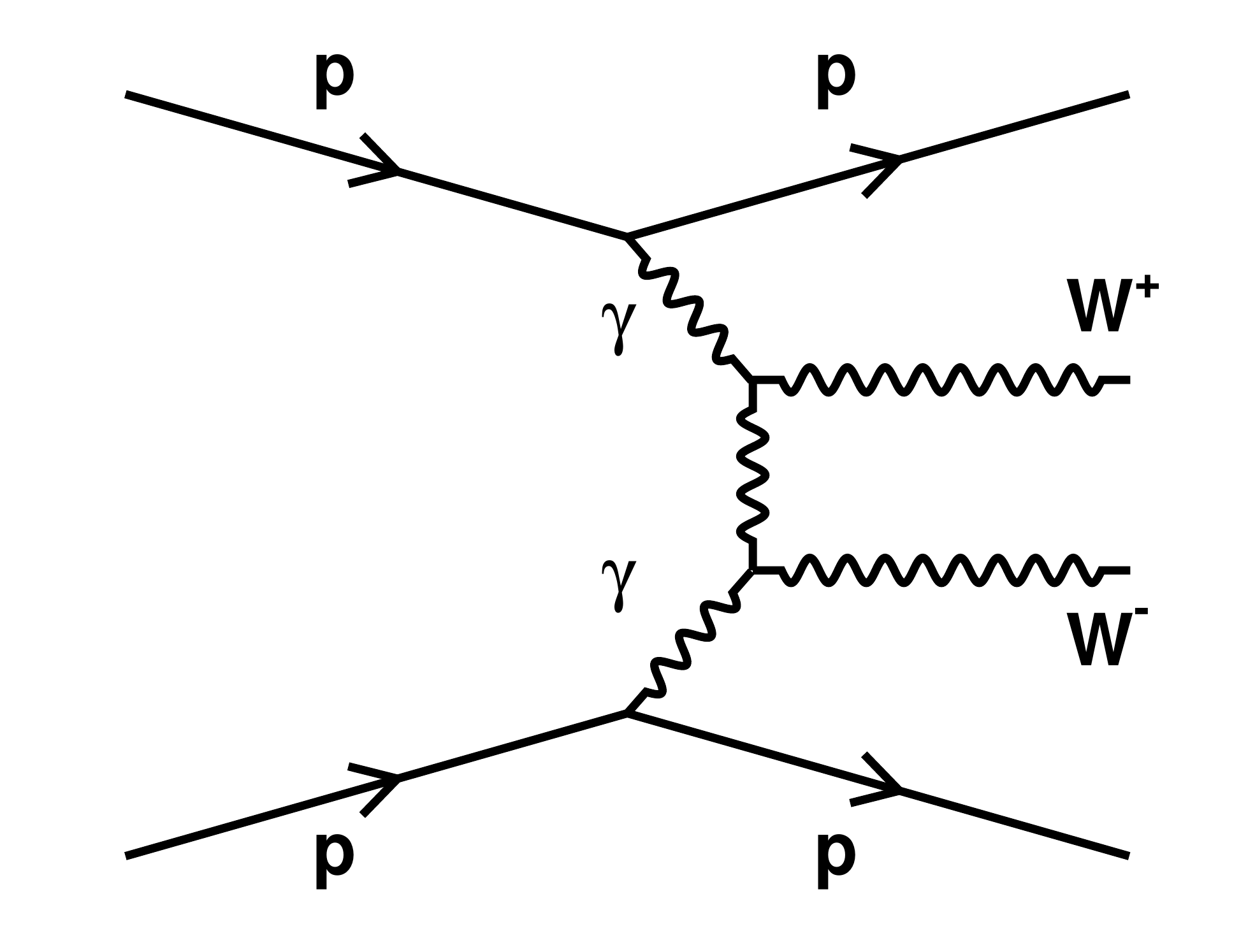}
\includegraphics[width=0.4\textwidth]{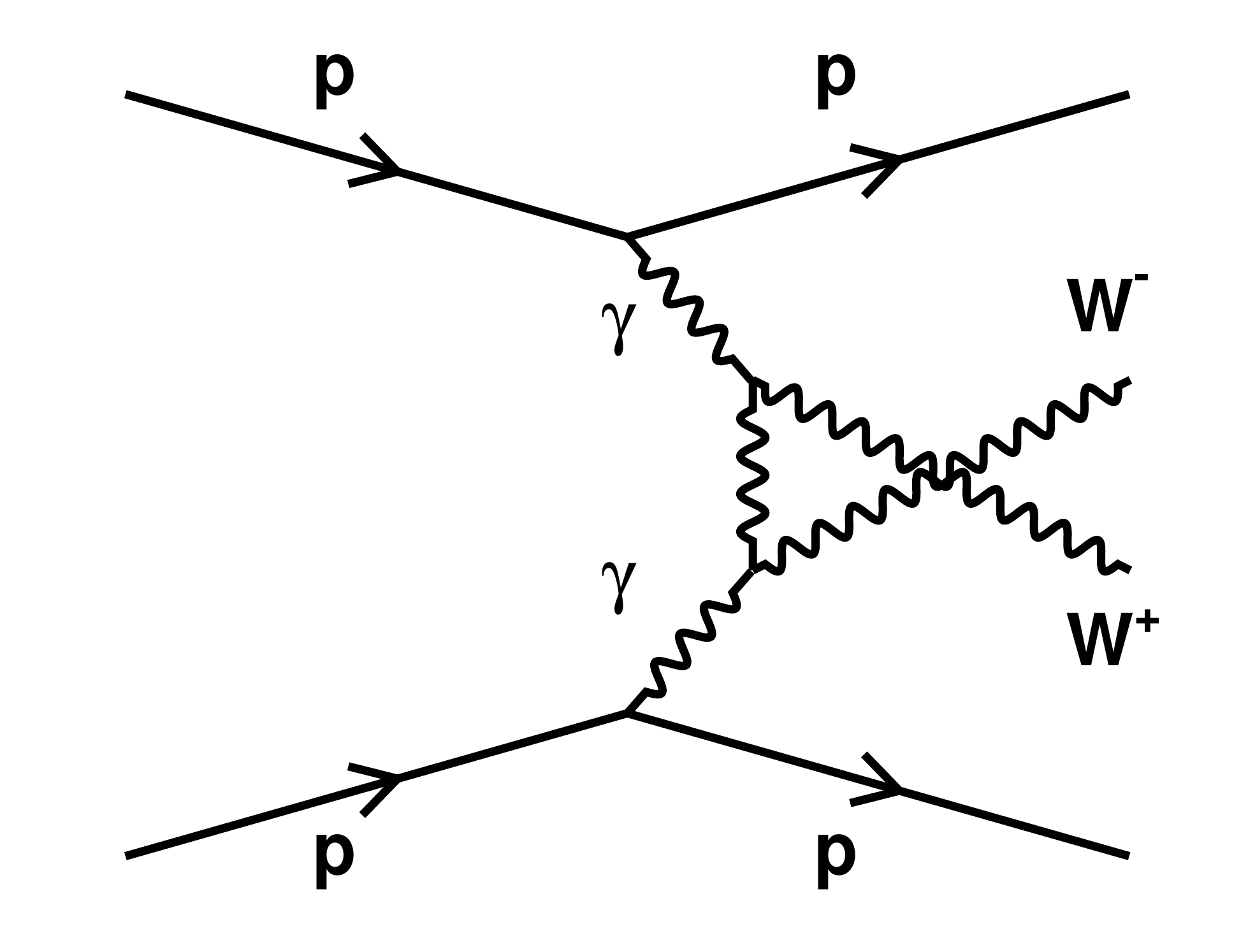}
\includegraphics[width=0.4\textwidth]{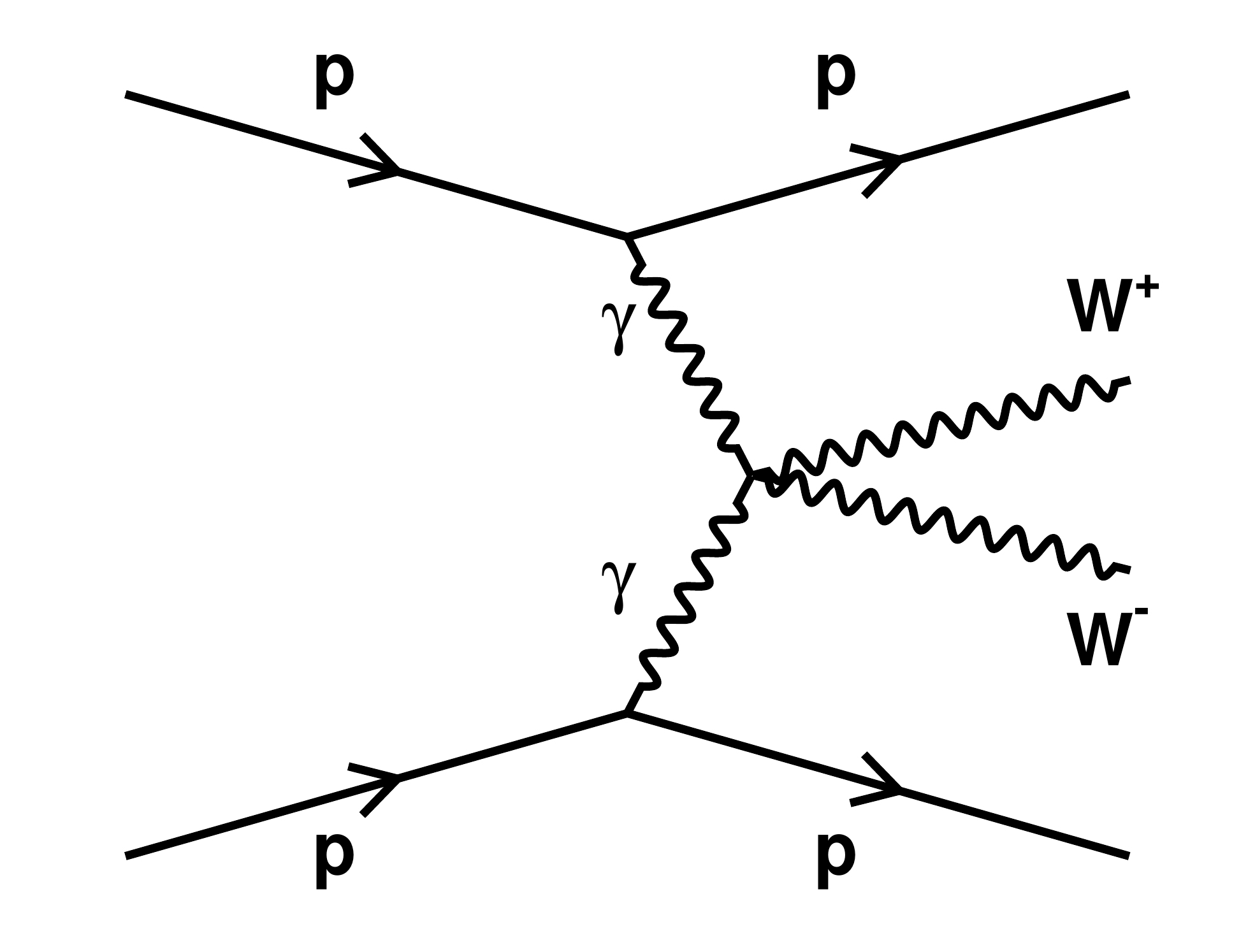}
\caption{Schematic diagrams of exclusive $\PGg\PGg \to \PW\PW$ production with intact protons according to the standard model (SM).}
\label{fig:feyndiagrams}
\end{figure}

A class of proton-proton scattering events at the CERN LHC exists in which the incoming protons radiate high-energy quasireal photons and remain intact, whereas only the two photons interact, as shown in Fig.~\ref{fig:feyndiagrams}. 
The protons lose a small fraction of their initial 6.5\TeV energy and are scattered at very small angles. In the CMS experiment, the scattered protons are measured with 
the Precision Proton Spectrometer (PPS), a set of near-beam detectors located on both sides of the central elements of the CMS detector at around 200\unit{m} from the interaction point (IP)~\cite{Albrow:2014lrm, Cms:2018het}. The trajectories of the scattered protons are bent by the LHC magnets between the interaction point and 
the PPS detectors, which makes it possible to measure their momenta. By reconstructing both protons, the center-of-mass 
energy of the two-photon collision is determined on an event-by-event basis.

In the present paper, a search is presented for the exclusive production of gauge boson pairs ($\PW\PW$ or $\PZ\PZ$) from $\PGg\PGg$ interactions using the PPS. The final state consists solely of the two bosons and the scattered protons, which are reconstructed using PPS. The fully hadronic decay modes of the $\PW$ and $\PZ$ bosons are studied.
When the gauge bosons are produced with a large boost, as expected in many scenarios beyond the standard model (BSM), decay products of each of the bosons are merged into a single large-area jet. 

The branching fraction of the fully hadronic channel for $\PW$ and $\PZ$ bosons is the highest (67--70\%), but this mode is not accessible without the proton detection because of the very large background from jet production in quantum chromodynamics (QCD) processes.
For exclusive production when the protons are measured, the kinematics of the final-state bosons can be independently reconstructed both with the central part of the CMS detector and with PPS. Therefore, for signal events that are correctly identified, the entire 13\TeV collision
energy can be reconstructed in the four-body $\Pp\PW\PW\Pp$ or $\Pp\PZ\PZ\Pp$ system.

Within the SM, quartic couplings involving two-photon production of charged ($\PWpm$) gauge bosons are allowed at tree level, as illustrated in Fig.~\ref{fig:feyndiagrams}.
Because of gauge invariance, the strength of these couplings is related to the triple gauge couplings, which enter through $t$- and $u$-channel
$\PGg\PGg\to \PW\PW$ production, and is fully specified in the SM. The SM cross sections for the $\PGg\PGg\to \PW\PW$ and $\PGg\PGg\to \PZ\PZ$ processes with both protons
intact are expected to be around 50\unit{fb} and $0.05$\unit{fb}, respectively, at $\sqrt{s} = 13\TeV$~\cite{Chapon:2009hh, Baldenegro:2020qut, Boonekamp:2011ky, Shao:2022cly}, and are concentrated at low values of the $\PV\PV$ invariant mass, 
$m(\PV\PV)$. Any significant signal over the prediction, particularly in the high $m(\PV\PV)$ tails where the expected SM
cross section is small, could indicate BSM physics. 

BSM effects on $\gamma\gamma\to\PV\PV$  processes are predicted in a variety of models, with both resonant and nonresonant 
signals~\cite{Pierzchala:2008xc,Chapon:2009hh,Maniatis:2008zz,Delgado:2014jda,Fichet:2013ola,Gupta:2011be,Delgado:2017cat,Delgado:2016rtd,Baldenegro:2020qut}. 
A common approach to quantify deviations from the SM with minimal assumptions involves anomalous quartic gauge couplings (AQGCs). Two formalisms commonly used in electroweak studies at the LHC exploit dimension-6 nonlinear operators, and 
dimension-8 linear operators~\cite{Belanger:1999aw,Belanger:1992qi,Eboli:2006wa, Eboli:2003nq, Eboli:2016kko}, where the terms linear and nonlinear refer to the 
realization of the broken electroweak symmetry. In the case of the dimension-6 operators, the most general formulation, which was used in the analysis of $\Pep\Pem$ 
collisions at LEP~\cite{OPAL:2004voc}, has four independent operators. The $a^{\PW}_{0}/\Lambda^2$ and $a^{\PW}_{C}/\Lambda^2$ operators, where $\Lambda$ is an energy scale characterizing 
new physics, will modify the $\PGg\PGg\to \PW\PW$ process if they have values different from the SM value of zero. Similarly, nonzero values of the $a^{\PZ}_{0}/\Lambda^2$ 
and $a^{\PZ}_{C}/\Lambda^2$ operators will modify the $\PGg\PGg\to \PZ\PZ$ process. The detailed definitions of these operators are reported in Refs.~\cite{Belanger:1999aw,Belanger:1992qi}.

The $\PGg\PGg\to \PW\PW$ process without the measurement of the outgoing protons was among several processes used to place the first LHC  
bounds on such AQGCs using 7 and 8\TeV data~\cite{ATLAS:2016lse,CMS:2013hdf,CMS:2016rtz,smp17,smp20}; constraints were also obtained from 1.96\TeV data at the Tevatron~\cite{D0:2013rce}. 
The cross section was also measured at 13\TeV, using the leptonic channel without proton detection~\cite{ATLAS:2020iwi}. Recently, strong constraints have been placed 
on AQGCs using a variety of scattering~\cite{CMS:2020gfh,CMS:2021gme,CMS:2020ypo,CMS:2017zmo,smp17,smp20} and triple boson production~\cite{CMS:2021jji} processes at 13\TeV.
The first limits on anomalous couplings using forward protons measured with PPS were obtained in searches for high-mass $\PGg\PGg \to \PGg\PGg$ scattering~\cite{TOTEM:2021kin}. 

\section{Experimental setup}

The central feature of the CMS apparatus is a superconducting solenoid of 6\unit{m} internal diameter, providing a magnetic field of 3.8\unit{T}. Within
the solenoid volume are a silicon pixel and strip tracker, a lead tungstate crystal electromagnetic calorimeter (ECAL), and a brass and scintillator hadron
calorimeter (HCAL), each composed of a barrel and two endcap sections. Forward calorimeters extend the pseudorapidity ($\eta$) coverage provided by the barrel and
endcap detectors. Muons are detected in gas-ionization chambers embedded in the steel flux-return yoke outside the solenoid.
A more detailed description of the CMS detector, together with a definition of the coordinate system used and the relevant kinematic variables, can be found
in Ref.~\cite{Chatrchyan:2008zzk}.

The Precision Proton Spectrometer is a system of near-beam tracking and timing detectors, located in Roman pots (RPs) at about 200\unit{m} from the CMS interaction
point~\cite{Albrow:2014lrm}. The Roman pots are movable near-beam devices that allow the detectors to be brought very close (within a few mm) to the beam without affecting the vacuum, 
beam stability, or other aspects of the accelerator operation. The PPS makes it possible to measure the 4-momentum of the scattered protons, along with their 
time-of-flight from the IP. The proton momenta are measured by two tracking stations on each arm of the spectrometer. 
The timing, not used for the present analysis, can be exploited to suppress background from other collisions in the same bunch crossing (pileup).

Events of interest are selected using a two-tiered trigger system. The first level (L1), composed of custom hardware processors, uses information from the
calorimeters and muon detectors to select events at a rate of around 100\unit{kHz} within a fixed latency of about 4\mus~\cite{Sirunyan:2020zal}. The second
level, known as the high-level trigger (HLT), consists of a farm of processors running a version of the full event reconstruction software optimized for fast
processing, and reduces the event rate to around 1\unit{kHz} before data storage~\cite{Khachatryan:2016bia}. The forward proton information from the PPS 
detector is available in the HLT, but is not used in the HLT selection.

\section{Data samples and simulation}

The data used for the present analysis were collected at $\sqrt{s} = 13\TeV$ in the years 2016--2018. The data are required to pass quality criteria for both PPS and 
other CMS subdetectors. In the case of PPS, these criteria require the Roman pots to be fully inserted in the data-taking position. Compared with CMS analyses that do not use forward protons, the requirement of having the RPs inserted removes a fraction of data at the 
beginning and end of each LHC fill. During the first year of operation (2016), a significant amount of time was also devoted to the commissioning of PPS, so the integrated luminosity available for the analysis for that year is only 9.9\fbinv. In the 2017 and 2018 samples, the integrated luminosities of the data meeting the PPS quality criteria are 37.2\fbinv and 52.9\fbinv, respectively, corresponding to nearly $90\%$ of the luminosity available for CMS analyses without forward protons. In total, the three years' data combined correspond to a luminosity of 100\fbinv~\cite{lum1,lum2,lum3}.

Signal events are simulated at leading order with the FPMC~\cite{Boonekamp:2011ky} Monte Carlo (MC) generator, for both the $\PGg\PGg \to \PW\PW$ and $\PGg\PGg \to \PZ\PZ$ channels. 
The dominant nonexclusive backgrounds from QCD multijet production are simulated at leading order with the \PYTHIA~8.205~\cite{Sjostrand:2014zea} MC generator with the CUETP8M1~\cite{Khachatryan:2015pea} tune for the 2016 samples, and \PYTHIA~8.230 with the CP5~\cite{Sirunyan:2019dfx} tune for the 2017--2018 
samples. Backgrounds arising from the production of a $\PW$ or $\PZ$ boson in association with jets are simulated at next to leading order (NLO) in QCD with \MGvATNLO~\cite{Madgraph:2014}. The background from pair production of top 
quarks is simulated at NLO in QCD with \POWHEG~\cite{Alioli:2010xa,Frixione:2007vw,Nason:2004rx}. The parton shower and hadronization for 
the $\PW+\text{jets}$, $\PZ+\text{jets}$, and \ttbar samples are performed with \PYTHIA.
The SM contribution to the $\PGg\PGg \to \PW\PW$ and $\PGg\PGg \to \PZ\PZ$ processes is expected to be negligible within the kinematic region considered in this analysis.

All of the generated signal and background samples are passed through a detailed \GEANTfour~\cite{GEANT4:2002zbu} simulation of the central part of the CMS detector, extending to $\abs{\eta} < 5$, and reconstructed in the same way as the data. For the signal samples, the forward protons are passed through a ``direct'' simulation~\cite{propaper}, which propagates the protons from the IP to the RP positions, simulates hits in the detector planes, and reconstructs the tracks and proton kinematics in the same way as done for the data. In the background samples, protons from pileup events are not simulated and their effect is estimated from the data, as described later. A realistic mix of beam crossing angles and aperture limitations along the beam line is used when simulating the protons.

\section{Event reconstruction}

Events are selected online by means of several jet triggers. These are based on either the highest transverse momentum (\pt) jet in the event, or the scalar sum of the \pt values of 
all jets, $H_{\mathrm{T}}$. In some cases, additional requirements on the ``trimmed'' mass~\cite{Krohn:2009th} of the jets are imposed, to select jet masses above 
30 or 50\GeV. The triggers are identical to those used in the analysis of Ref.~\cite{CMS:2019drn}, where a more detailed description is reported.

The primary vertex (PV) is taken to be the vertex corresponding to the hardest scattering in the event, evaluated using tracking information alone, as described in Section 9.4.1 of Ref.~\cite{CMS-TDR-15-02}.

The particle-flow algorithm~\cite{CMS:2017yfk} reconstructs and identifies each individual particle in an event, with an optimized combination of information from the various 
elements of the CMS detector. The energy of photons is obtained from the ECAL measurement. The energy of electrons is determined from a combination of the electron momentum at the 
primary interaction vertex as determined by the tracker, the energy of the corresponding ECAL cluster, and the energy sum of all bremsstrahlung photons spatially compatible with 
originating from the electron track. The energy of muons is obtained from the curvature of the corresponding track. The energy of charged hadrons is determined from a combination 
of their momentum measured in the tracker and the matching ECAL and HCAL energy deposits, corrected for the response function of the calorimeters to hadronic showers. Finally, 
the energy of neutral hadrons is obtained from the corresponding corrected ECAL and HCAL energies.

For each event, hadronic jets are clustered from these reconstructed particles using the infrared and collinear safe anti-\kt algorithm~\cite{Cacciari:2008gp}, 
as implemented in the \FASTJET package~\cite{Cacciari:2011ma}, with 
a distance parameter of 0.8. Jet momentum is determined as the vectorial sum of all particle momenta in the jet, and is found from simulation to be, on average, within 
5 to 10\% of the true momentum over the whole \pt spectrum and detector acceptance. Additional proton-proton interactions from pileup 
can contribute with additional tracks and calorimetric energy depositions to the jet momentum. To mitigate this effect, charged particles identified to be originating from pileup vertices 
are discarded and an offset correction is applied to remove any remaining contributions. Jet energy corrections are derived from simulation to bring the measured response of jets 
to that of particle-level jets on average. In situ measurements of the momentum balance in dijet, $\PGg + \text{jet}$, $\PZ + \text{jet}$, and multijet events are used to 
account for any residual differences in the jet energy scale between data and simulation~\cite{CMS:2016lmd}. The jet energy resolution amounts typically to 15--20\% 
at 30\GeV, 10\% at 100\GeV, and 5\% at 1\TeV~\cite{CMS:2016lmd}. Standard CMS quality criteria are applied to each jet to remove jets potentially dominated by spurious contributions from various subdetector components or reconstruction failures.

Hadronic decays of $\PW$ and $\PZ$ bosons are identified using the ratio between 2-subjettiness and 1-subjettiness~\cite{Thaler:2010tr},
$\tau_{21}=\tau_{2}/\tau_{1}$, and the jet mass $m(\Pj)$ after applying a ``pruning'' algorithm to reduce the contributions of soft gluon radiation and
pileup~\cite{Ellis:2009me,Khachatryan:2014vla}. The pruning algorithm is a procedure to repeat the initial jet clustering, removing the softest and largest angle components at 
each step. For each pair of jet constituents the hardness $z$ is defined as the ratio of the minimum $\pt$ of the pair to the $\pt$ of their combination. The constituent with 
the lower $\pt$ is removed if the hardness is below the threshold $z_\text{c} = 0.1$, and if the angular distance relative to the axis of the combination of the two constituents 
is larger than $D_\text{c} = 0.5 m/\pt$ (where $m$ and $\pt$ are  the mass and transverse momentum of the original jet). If neither constituent is removed, they are merged 
into a new constituent. This procedure is repeated iteratively until all constituents have either been removed or merged. The pruned mass obtained from this final jet is then 
used in the remainder of the analysis. 

The $\tau_{N}$ subjettiness variables are defined by the $\pt$ weighted $\Delta R$ distance between each jet constituent particle and the nearest subjet. When all constituent 
particles are close to a correctly identified subjet, $\tau_{N}$ will have a small value indicating the jet is consistent with containing ${\leq}N$ subjets. The ratio 
$\tau_{21}=\tau_{2}/\tau_{1}$ will then be smaller if the jet is more consistent with containing two subjets, as expected for a $\PW$ or $\PZ$ boson, instead of one subjet, as expected for a quark or gluon jet. Since $\tau_{21}$ may have an undesirable correlation with the jet mass and $\pt$, a ``Designed Decorrelated Taggers'' (DDT) 
approach~\cite{Dolen:2016kst,CMS:2019drn} was developed to reduce this. Simulated samples are used to extract the slope $M$ of the mean value of $\tau_{21}$ vs. $\rho'$, 
where $\rho' = \ln(m^{2}_\text{jet}/ (\pt \times 1\GeV))$. The slope $M$ is then used to define the decorrelated subjettiness 
variable $\tau_{21}^{\mathrm{DDT}} = \tau_{21} - M \, \rho'$.

Forward protons are reconstructed using a ``multi-RP'' algorithm, which combines tracks reconstructed in both of the tracking Roman pots in each arm of PPS.
The lever arm between the two RPs allows the reconstruction of the proton scattering angles $\theta^*$ (where the superscript ``*'' indicates an angle  
defined at the IP) and proton fractional momentum loss $\xi$:
\begin{linenomath}
\begin{equation}
\xi = (p_{\text{nom}} - p) / p_{\text{nom}} ,
\label{eq:introduction-xi}    
\end{equation}
\end{linenomath}
where $p_{\text{nom}}$ and $p$ are the nominal beam momentum and the scattered proton momentum, respectively. PPS provides a $\xi$ resolution of ${\leq}5\%$ over the full acceptance~\cite{propaper}. Additional selections are applied to ensure that the protons are within the acceptance
of the sensors, and within the aperture constraints defined by the LHC collimators. The LHC conditions and the PPS tracking detector configuration were 
different for each of the three data-taking years. For this reason, the 2016, 2017, and 2018 samples are 
analyzed separately, and the results are combined at the end. Further details of the proton reconstruction, including the determination of the 
beam optics and the detector alignment, are presented in~\cite{propaper, Nemes:2256433, Kaspar:2256296}.

\section{Event selection}

Events are selected based on the properties of the jets, the protons, and the correlation between protons and jets. 

\subsection{Jet selection}

\begin{figure}[h!t]
\centering
\includegraphics[width=0.42\textwidth]{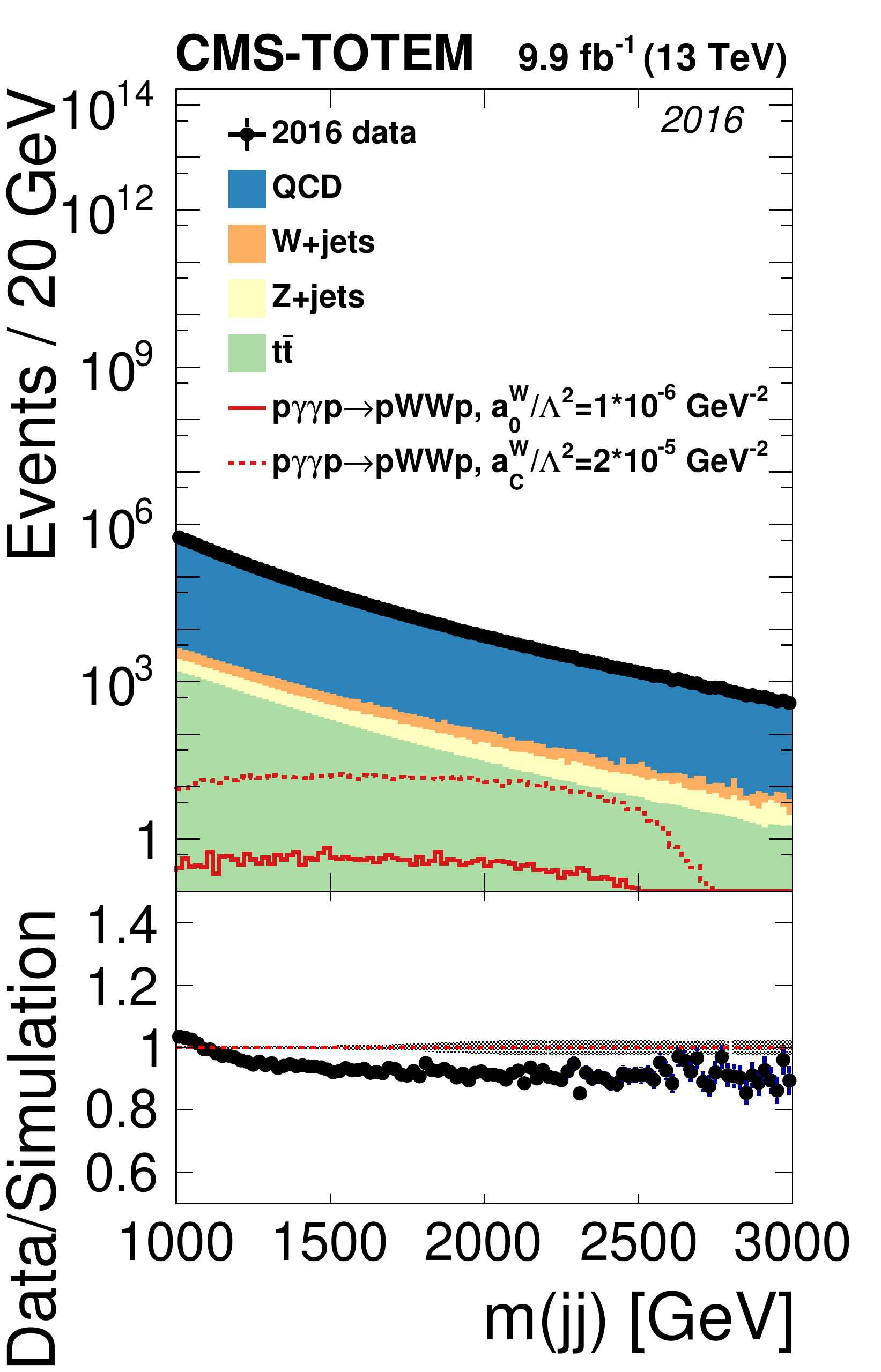}
\includegraphics[width=0.42\textwidth]{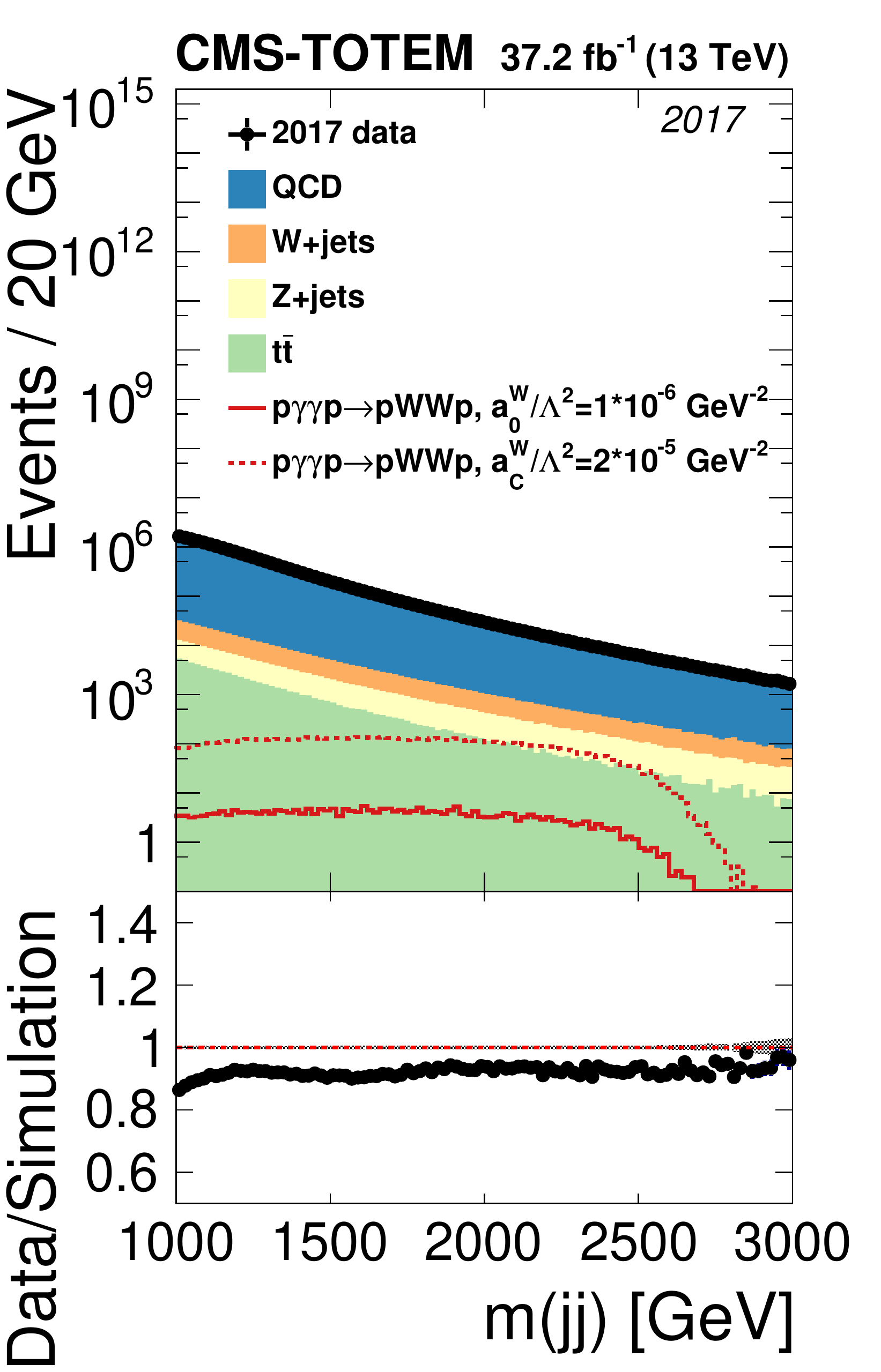}
\includegraphics[width=0.42\textwidth]{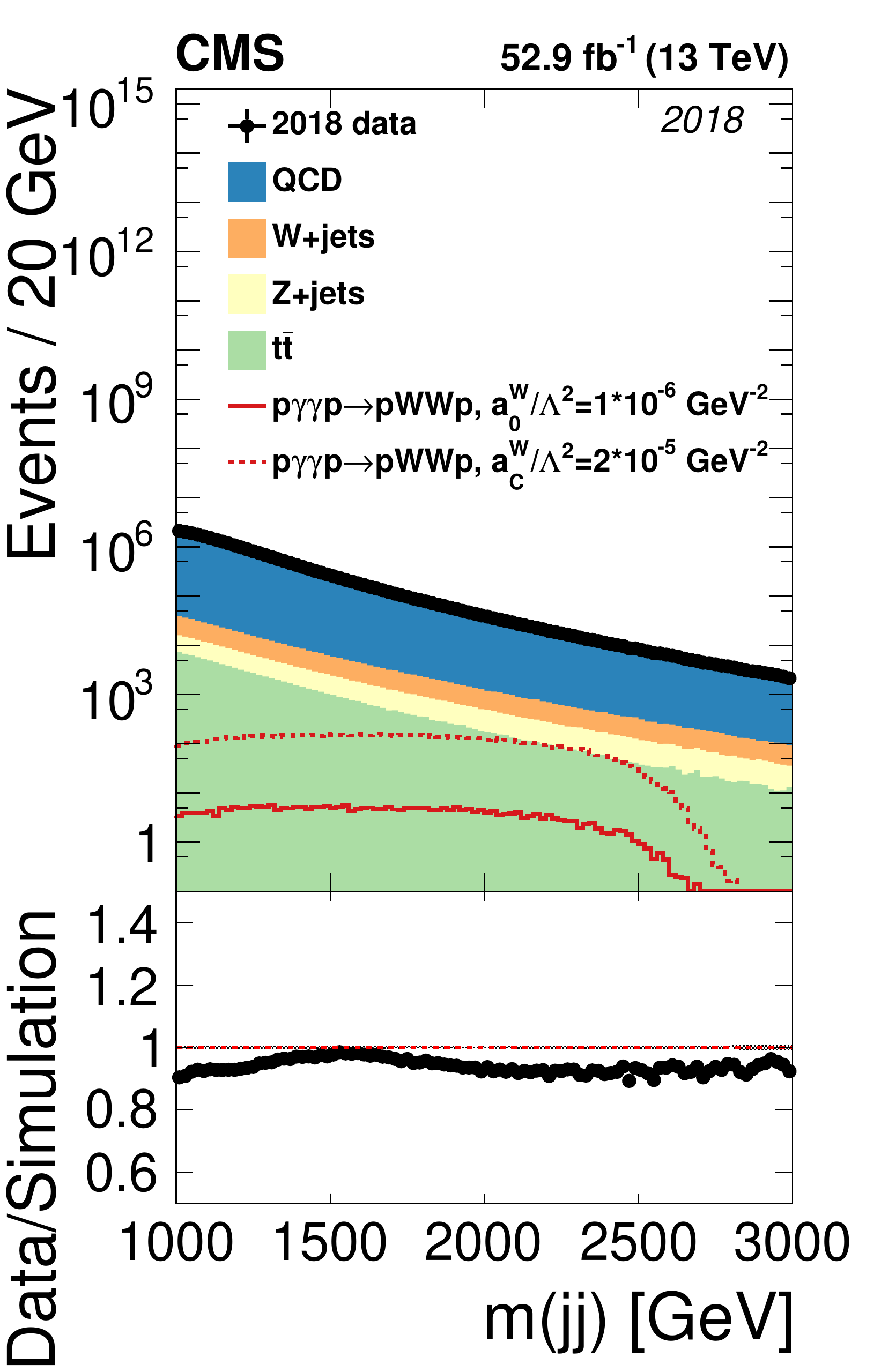}
\caption{Dijet invariant mass spectra in data and simulation, for the years 2016 (upper left), 2017 (upper right), and 2018 (lower). The distributions of number of events show data compared with the stacked background predictions from simulation, with the corresponding ratios of data to the sum of simulated backgrounds, shown below them. The plots are shown at the preselection level, with no requirements on the protons, jet substructure, or dijet balance. Examples of simulated signals are shown for protons generated in the range of $0.01 \le \xi \le 0.20$. Only statistical uncertainties (dashed grey bands) are shown.}
\label{fig:mjjpreselection}
\end{figure}

The jets are first required to have $\pt > 200\GeV$, $\abs{\eta} < 2.5$, and, if more than two pass these criteria, the two with the highest \pt are chosen.
In the following, these are labeled $\Pj1$ and $\Pj2$, corresponding to the jet with the highest and second-highest \pt, respectively. 
The pair of jets is required to have a dijet invariant mass $m(\Pj\Pj) > 1126\GeV$, where the efficiency of the trigger is ${>}99\%$ relative to the offline 
selection~\cite{CMS:2019drn}. The two jets are further required to have a pseudorapidity difference of $\abs{\Delta \eta} < 1.3$ to
reduce the QCD multijets background. Figure~\ref{fig:mjjpreselection} shows the dijet invariant mass distribution in data and simulation for each of the three years of data taking in the range of interest.

To enhance the exclusive production, the jets are required to be balanced in azimuthal angle and transverse momentum. This is implemented by requiring the 
acoplanarity ($a = \abs{1-\abs{(\phi_{\Pj1}-\phi_{\Pj2})}/\pi}$) is $< 0.01$, and the ratio ($\pt(\Pj1)/\pt(\Pj2)$) is $< 1.3$. The $\phi$ resolution for 
single jets in the region of interest is approximately $0.004$ radians, and the resulting efficiency of the $a < 0.01$ requirement is $99\%$ for 
simulated AQGC signal events. 

Finally, the jet substructure properties are exploited to enrich the sample with merged jets from boosted $\PW$ or $\PZ$ decays. The pruned mass of the jets is required to 
be between $60$ and $107\GeV$, \ie, compatible with the $\PW$ or $\PZ$ mass. The constraint retains almost all signal events. The $\tau_{21}^{\mathrm{DDT}}$ discriminator 
must be smaller than 0.75, compatible with that expected for two quark jets originating from one boson decay and merged into a single high-momentum jet. 
From simulation, this requirement retains approximately $85\%$ of anomalous $\PGg\PGg \to \PW\PW$ events passing all other jet selection criteria, while rejecting 
${\approx}65\%$ of high mass QCD multijet background events. 

\subsection{W and Z selection}

\begin{figure}[ht]
\centering
\includegraphics[width=0.85\textwidth]{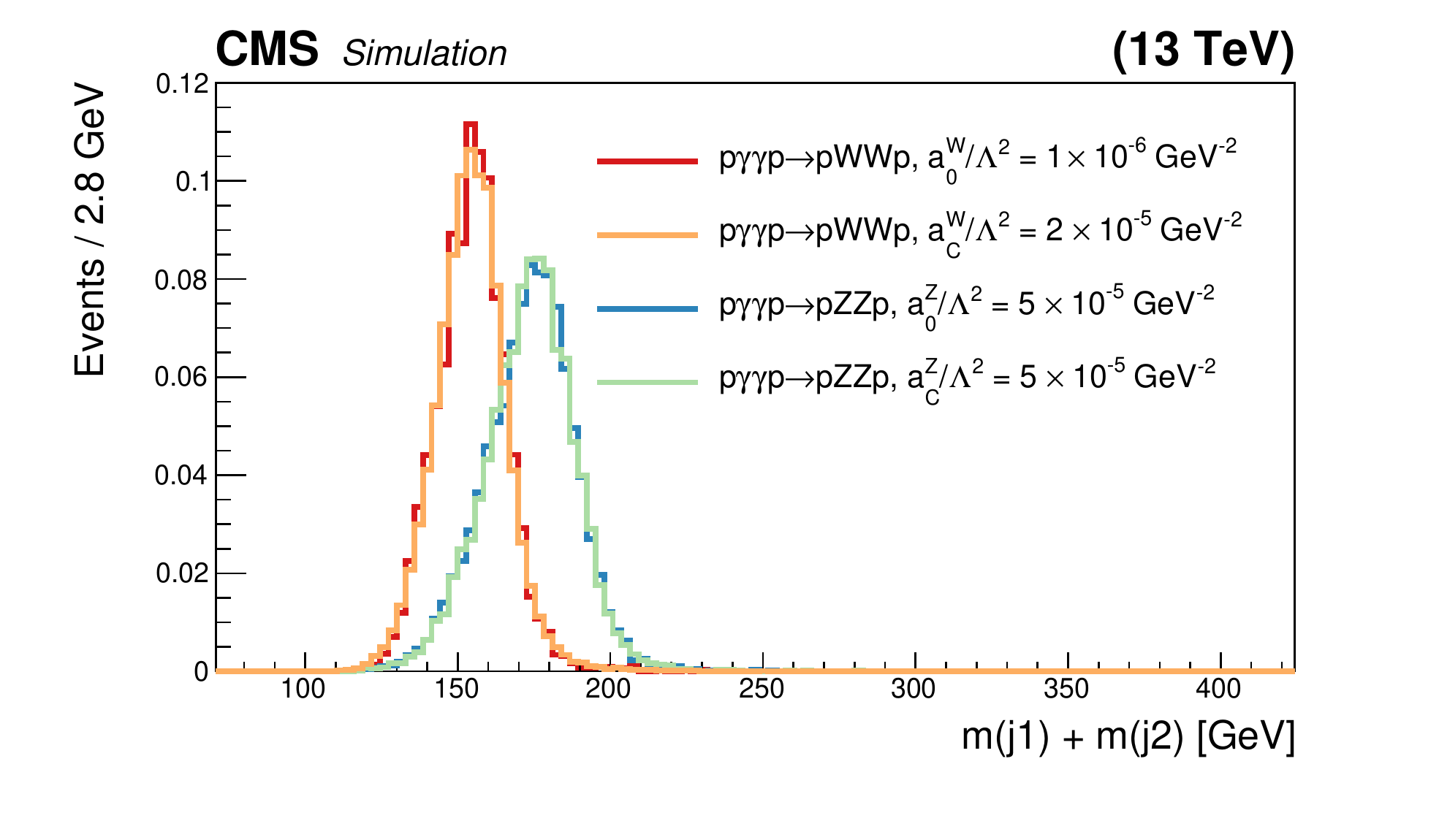}
\caption{Event distribution as a function of the discriminant described in the text for simulated $\PW\PW$ and $\PZ\PZ$ signal events.}
\label{fig:sigsep}
\end{figure}

The reconstructed pruned mass of the jets discriminates between the $\PW\PW$ and $\PZ\PZ$ final states, due to the mass difference between
the $\PW$ and $\PZ$ bosons. To reduce the two-dimensional space of the pruned jet masses (denoted by $m(\Pj1)$ and $m(\Pj2)$)
into a single variable, a simple discriminant based on the sum of $m(\Pj1) + m(\Pj2)$ is used.
The result is shown for simulated signal events in Fig.~\ref{fig:sigsep}, with two distinct distributions visible, corresponding to the
$\PW\PW$ and $\PZ\PZ$ samples. The resolution of the resulting variable is ${\approx}8\GeV$ for the $\PW\PW$ and 10\GeV for the $\PZ\PZ$ sample, correspondingly. 
The optimal value to separate the two channels is then chosen by first treating one of the channels as signal and the other as background, and then inverting their roles. A single constraint is chosen that maximizes the sum (S($\PW\PW$)+S($\PZ\PZ$))/(B($\PW\PW$)+B($\PZ\PZ$)), where S indicates that the $\PW\PW$ or $\PZ\PZ$ candidates are reconstructed in the correct mass region (i.e. the one containing the peak of the $\PW\PW$ or $\PZ\PZ$ distribution, respectively), and B indicates that they are reconstructed outside that region. 
A value of $m(\Pj1)+m(\Pj2) = 166.6\GeV$ provides the best separation between $\PW\PW$ and $\PZ\PZ$ events; this retains more than 80\% of signal for each channel, and corresponds to a signal to background ratio of ${\approx}4.8$, with no significant dependence on the anomalous coupling values assumed.

\subsection{Proton selection}

Only protons with a $\xi$ value greater than 0.05 are retained for the analysis to avoid the region of large radiation-induced inefficiencies of the PPS detectors near the beam.
For signal events, the $\xi > 0.05$ requirement is less restrictive than the $m(\Pj\Pj) > 1126\GeV$ condition imposed on the jets. For background events, where the protons are uncorrelated with the jets, the 
$\xi > 0.05$ and $m(\Pj\Pj) > 1126\GeV$ requirements are independent. The upper bounds imposed by the collimators 
are different for each of the two arms of the spectrometer, each data-taking period, and each beam crossing angle. The exact upper $\xi$ requirements therefore vary 
with the data-taking period, generally lying within the range $0.12 \le \xi \le 0.20$. The upper mass limit of the centrally produced system (dijet in this analyis) 
correspondingly varies between $1.55$ and $2.01\TeV$ (for a proton scattering angle of $\theta^{*}=0$). 

In the 2016 and 2017 data samples, only one proton per event can be reconstructed in each arm of the spectrometer. When multiple protons are present, the pattern recognition 
is unable to find a unique solution, and no protons are reconstructed in that arm. With improved detectors in 2018, multiple protons can be 
reconstructed. This leads to an increase in both the expected number of observed signal events $S$, and the expected number of observed background events $B$.
In terms of the expected significance, estimated as $S/\sqrt{B}$, the use of up to three tracks in each arm leads to ${\approx}7\%$ improvement over using only events with one 
track. No additional improvement is seen by using events with more than three tracks, where the effect of showers and noncollision backgrounds becomes significant.
Therefore, only events with up to three protons per arm are used, and the proton with the largest value of $\xi$ is chosen for the analysis. 
This is motivated because the anomalous signal cross section is rising with $\xi$, whereas the SM backgrounds are generally decreasing with $\xi$. When a simulated signal proton 
is within the acceptance of the detectors, this choice results in the correct proton being selected $63$ to $77\%$ of the time, depending on the arm of the spectrometer. In 
simulated signal events where both protons are within the acceptance, the probability of correctly selecting at least one of the two is $92\%$. The 
use of multiple-proton events leads to both a significantly larger signal efficiency and a larger combinatorial background in the 2018 data, compared with the 2016 or 2017 data. 

\subsection{Proton-jet matching and signal region}

The matching between the proton and jet kinematics for exclusive signal is based on the variables $1 - m(\VV)/m(\Pp\Pp)$ and $y(\Pp\Pp) - y(\VV)$. Here $m(\VV)$ and $y(\VV)$ represent the invariant mass and rapidity 
of the $\PW\PW$ or $\PZ\PZ$ system, as reconstructed from the merged jets. The variables $m(\Pp\Pp)$ and $y(\Pp\Pp)$ are the expected invariant mass and rapidity of the central system, calculated from 
the proton information:
\begin{linenomath}
\begin{equation}
m(\Pp\Pp) = \sqrt{s} \sqrt{\xi_{\Pp 1}\xi_{\Pp 2}}, \qquad y(\Pp\Pp) = -\frac{1}{2} \ln\left(\frac{\xi_{\Pp 1}}{\xi_{\Pp 2}}\right).
\end{equation}
\end{linenomath}

\begin{figure}[ht]
\centering
\includegraphics[width=0.8\textwidth]{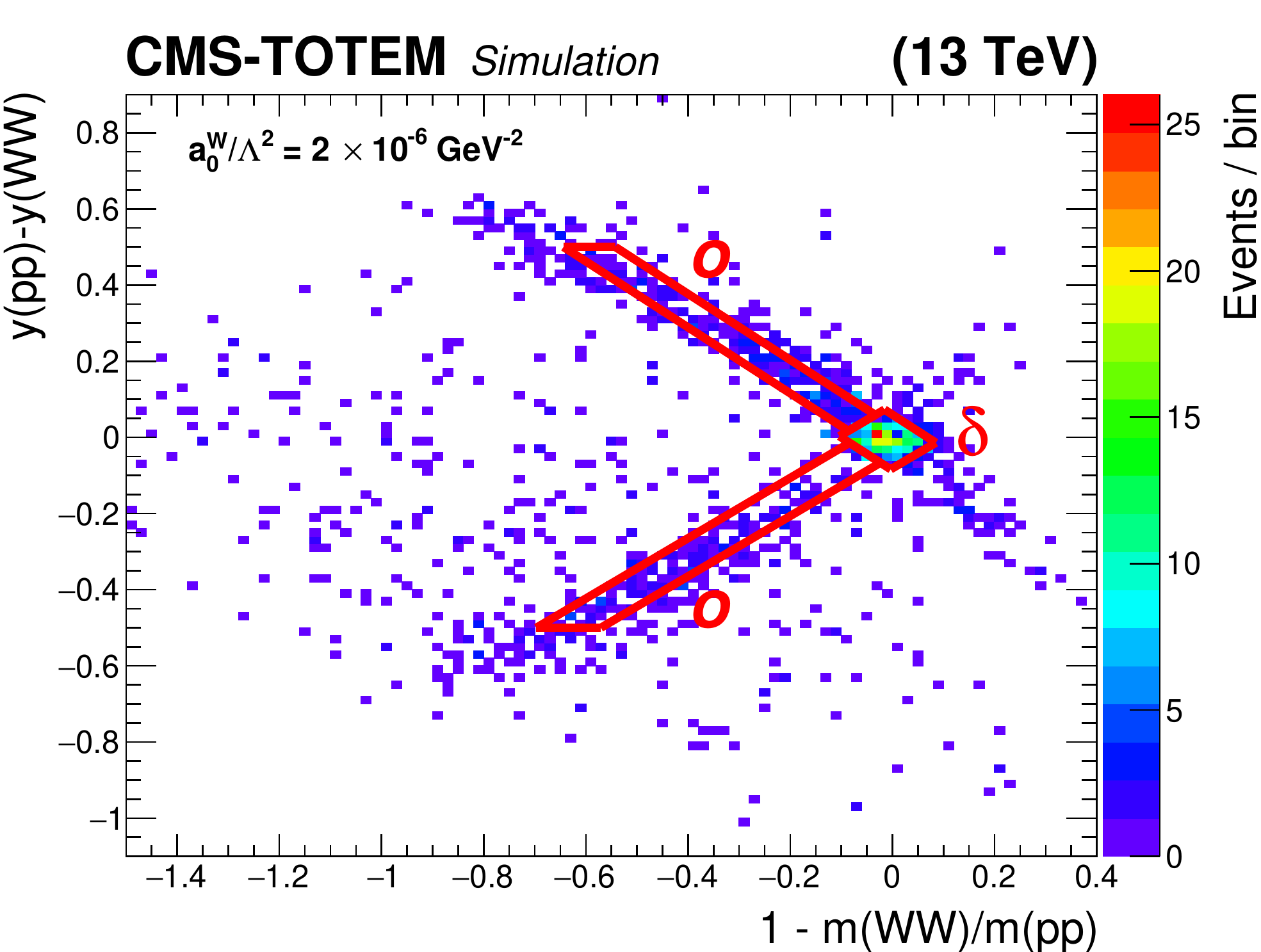}
\caption{Matching between the jets and the protons, in invariant mass and rapidity, for simulated $\PW\PW$ exclusive signal events. The red diamond-shaped area near the origin (signal region $\delta$) corresponds to the case where both protons are correctly matched to the jets. The areas within the red diagonal bands (signal region $o$) correspond to the case where one proton is correctly matched, and the second proton originates from a pileup interaction.}
  \label{fig:massrapiditymatchsignalmc}
\end{figure}

The resolution of $m(\Pp\Pp)$ is estimated to be 2.3\% from simulation. Two signal regions are defined by comparing the invariant mass and rapidity of the $\PW\PW$ or $\PZ\PZ$ system obtained from the jets with the same quantities inferred from the two protons. 
A diamond-shaped area in the $y(\Pp\Pp) - y(\VV)$ vs. $1 - m(\VV)/m(\Pp\Pp)$ plane, centered around zero, contains the bulk of the signal when both protons are correctly associated to the 
jets (``region $\delta$''). In case one of the signal protons is missed and a pileup proton is used instead, the events tend to fall in one of the two diagonal bands 
of Fig.~\ref{fig:massrapiditymatchsignalmc}. A second signal region (``region $o$'') is therefore defined based on these bands. The dimensions of the two regions are optimized to provide 
the best expected signal significance, estimated as $S/\sqrt{B}$. Of the simulated signal events passing all other selections, 
typically between 19 and 25\% are contained in region $\delta$, and a similar fraction in region $o$. The area with $\abs{1 - m(\VV)/m(\Pp\Pp)} < 1.0$ and $\abs{y(\Pp\Pp) - y(\VV)} < 0.5$, 
encompassing both signal regions, remained blinded and was not examined until the selection criteria and background estimation methods were fixed. 

\section{Background estimation}

The background is mainly due to jets produced in one $\Pp\Pp$ collision combined with unrelated protons from pileup interactions in the same bunch crossing.
The largest source of jets is QCD multijet production, with smaller contributions from $\PW$ or $\PZ$ bosons in association with jets, and $\ttbar$ production.
The protons predominantly arise from diffractive pileup interactions, which are not expected to be well modeled by simulations. For this reason, we rely mainly
on data to estimate the background.

The nominal background estimate is derived by inverting the dijet acoplanarity requirements, and/or the dijet-proton matching requirements, to define three
independent sideband regions. The first (region B) is defined by acoplanarity $a > 0.01$, with the signal region selection applied in $\abs{1 - m(\PW\PW)/m(\Pp\Pp)}$ and $\abs{y(\Pp\Pp)-y(\PW\PW)}$. 
The second region (region C) has $a < 0.01$, and $\abs{1 - m(\PW\PW)/m(\Pp\Pp)} > 1.0$ or $\abs{y(\Pp\Pp)-y(\PW\PW)} > 0.5$. The third (region D) is defined by 
 $a > 0.01$, and $\abs{1 - m(\PW\PW)/m(\Pp\Pp)} > 1.0$ or ${y(\Pp\Pp)-y(\PW\PW)} > 0.5$. If the numbers of events in each of these regions are $N_{\mathrm{B}}$, $N_{\mathrm{C}}$, and $N_{\mathrm{D}}$, respectively, 
the expected number of events in the signal region is then $N_{\mathrm{A}} = (N_{\mathrm{B}}\,N_{\mathrm{C}}) / N_{\mathrm{D}}$.

\begin{figure}[ht]
\centering
\includegraphics[width=0.49\textwidth]{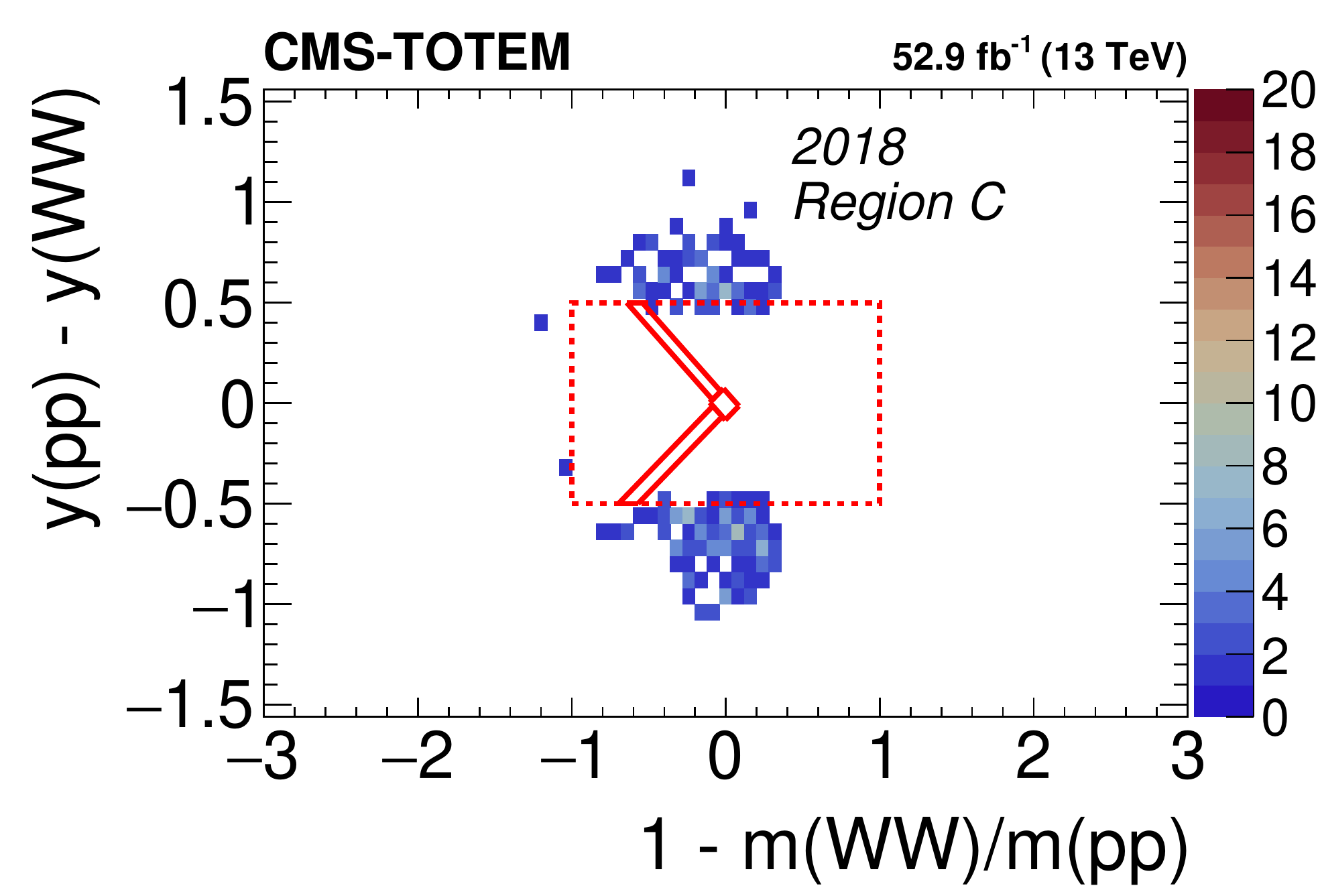}
\includegraphics[width=0.49\textwidth]{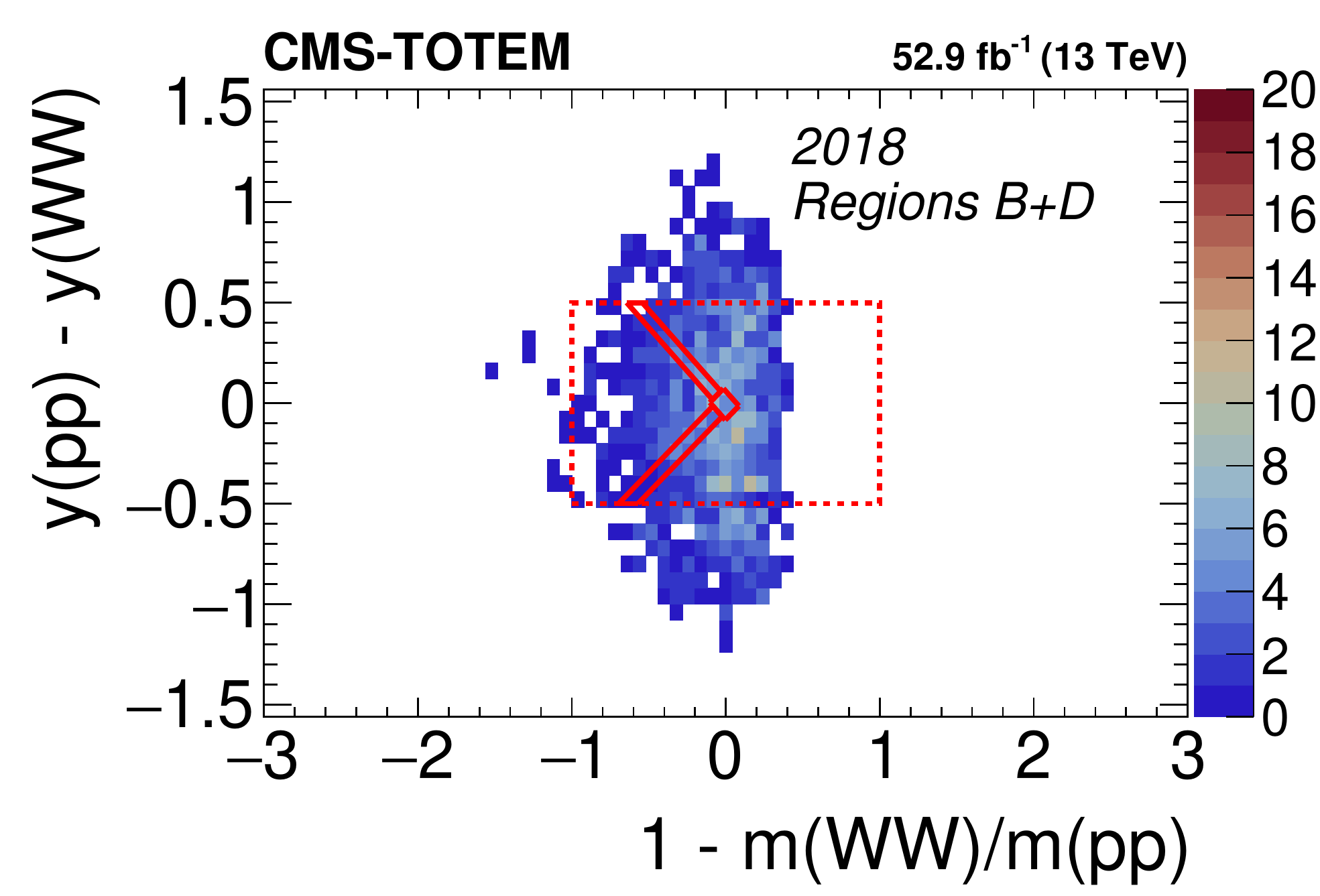}
\caption{Distribution of the 2018 data in the $y(\Pp\Pp) - y(\VV)$ vs $1 - m(\VV)/m(\Pp\Pp)$ plane in the $\PW\PW$ mass region. On the left, the sample with  
all selections applied is shown, except that the region inside the dashed lines remains blinded. On the right, the region with the acoplanarity requirement inverted 
is shown. The solid lines indicate the same signal regions as shown in Fig.~\ref{fig:massrapiditymatchsignalmc}. In the right plot the area 
inside the solid lines corresponds to ``region B'', while the area outside the dashed lines corresponds to ``region D''. The color scale on the $z$ axis indicates 
the number of events in each bin.}
\label{fig:massrapiditymatchsidebands}
\end{figure}

\begin{figure}[ht]
\centering
\includegraphics[width=0.49\textwidth]{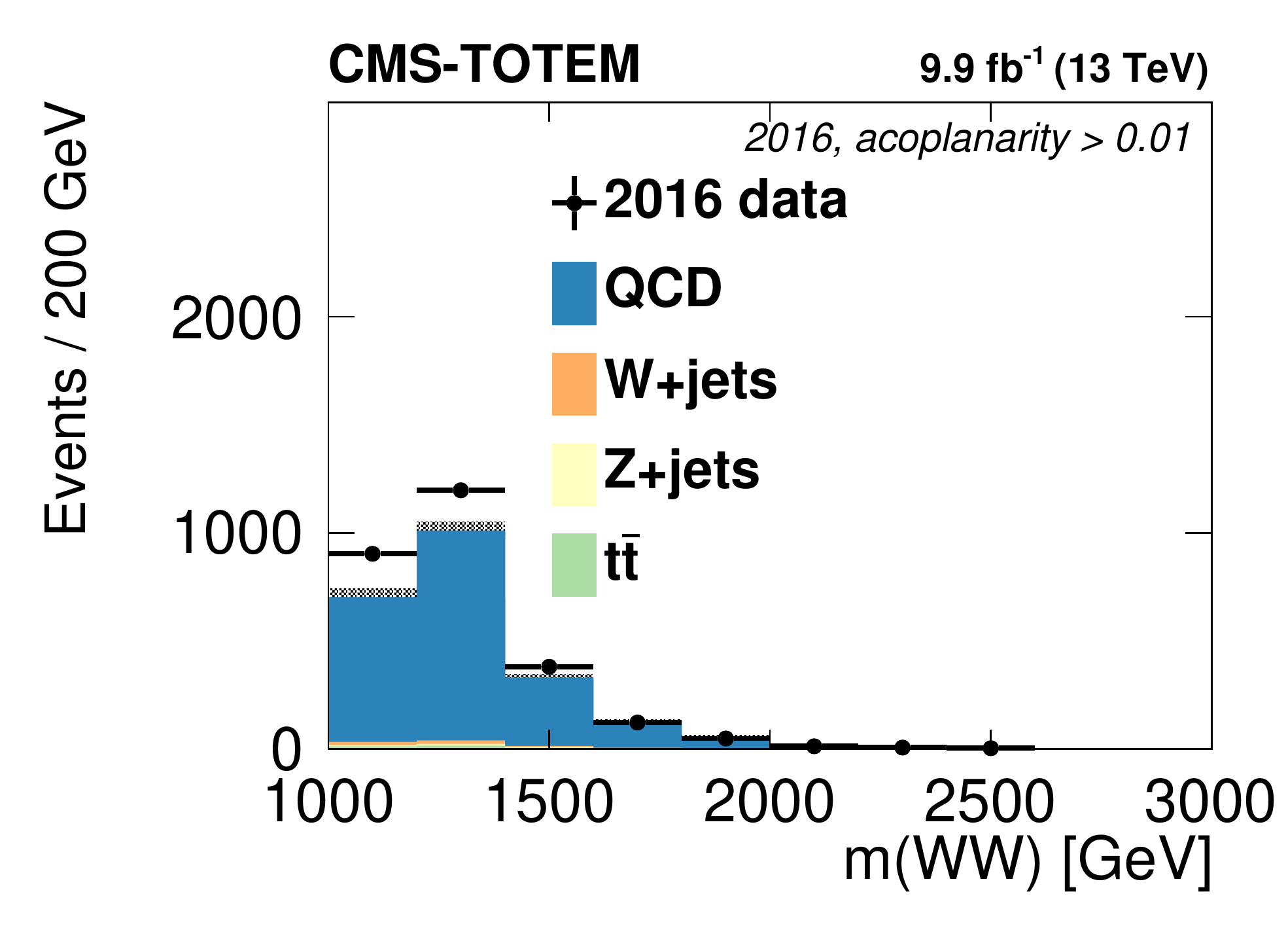}
\includegraphics[width=0.49\textwidth]{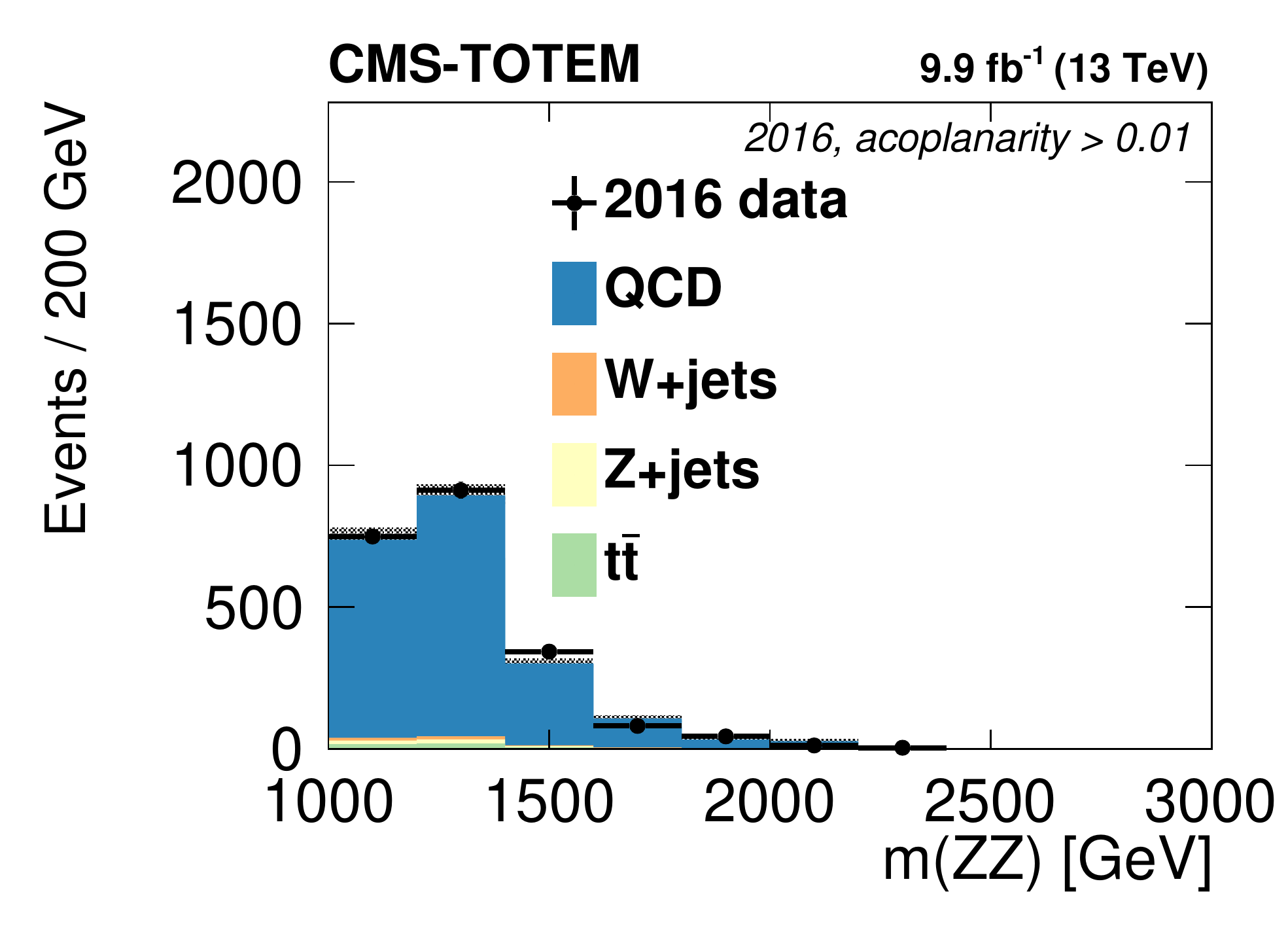}

\includegraphics[width=0.49\textwidth]{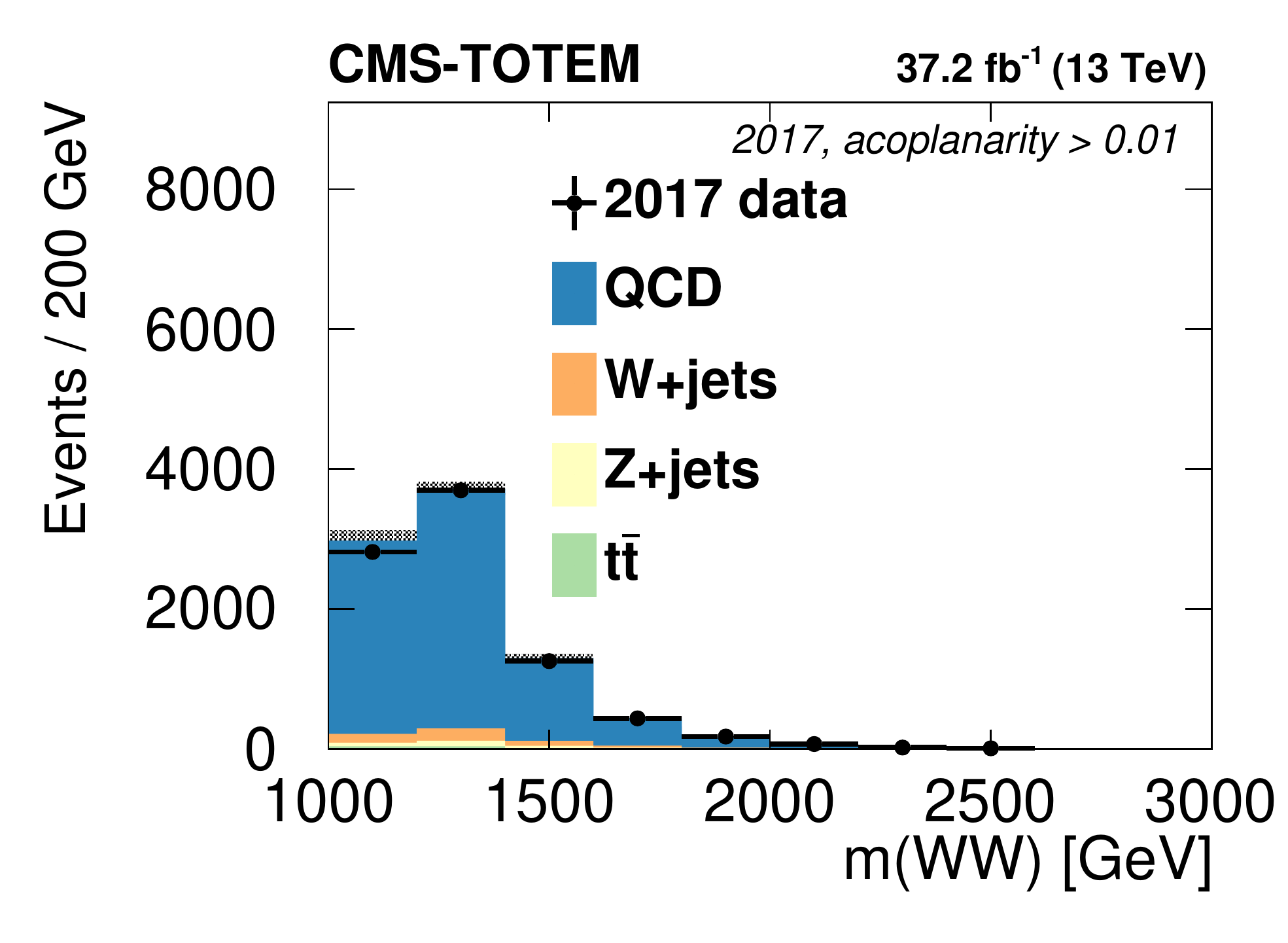}
\includegraphics[width=0.49\textwidth]{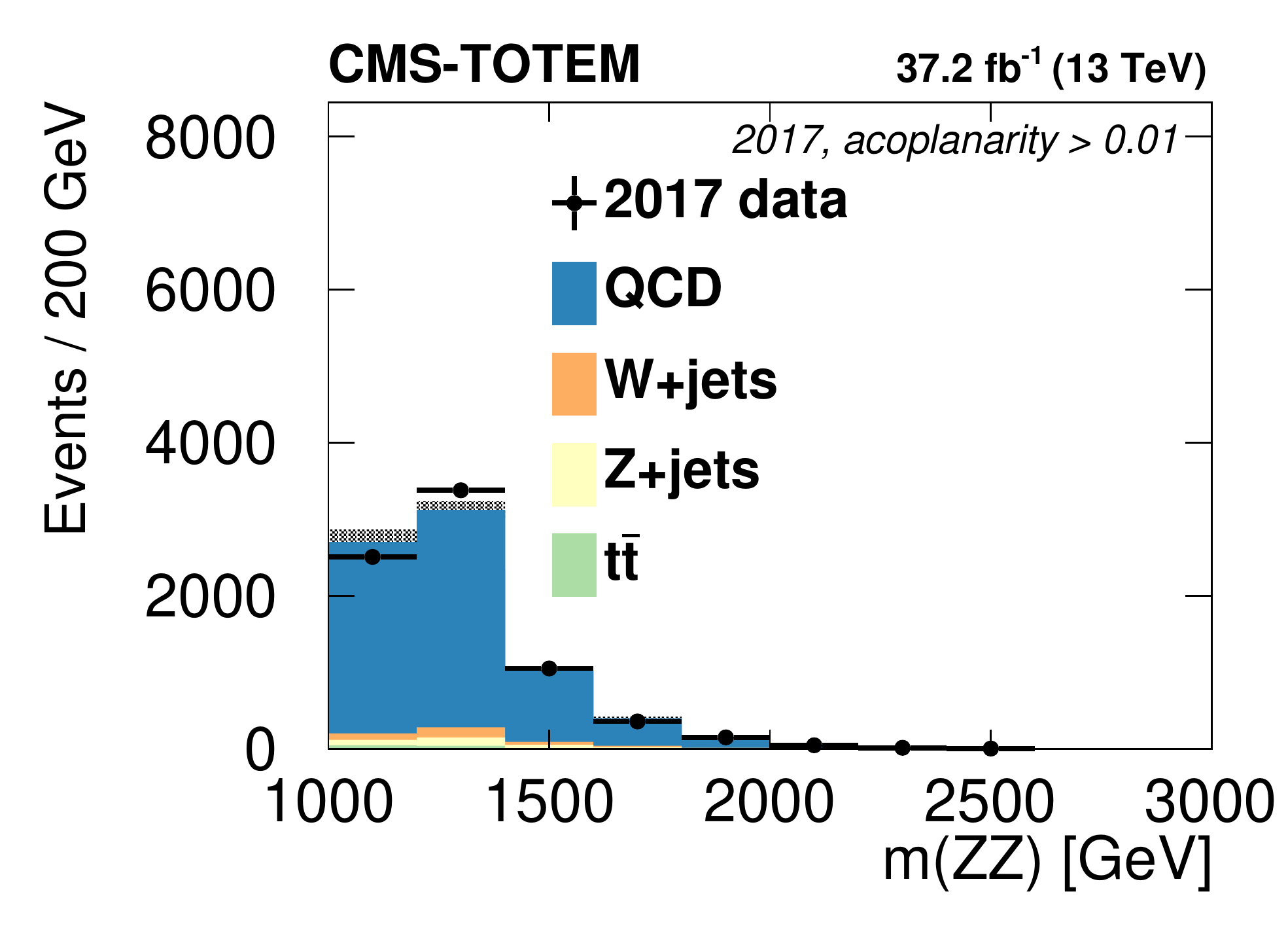}

\includegraphics[width=0.49\textwidth]{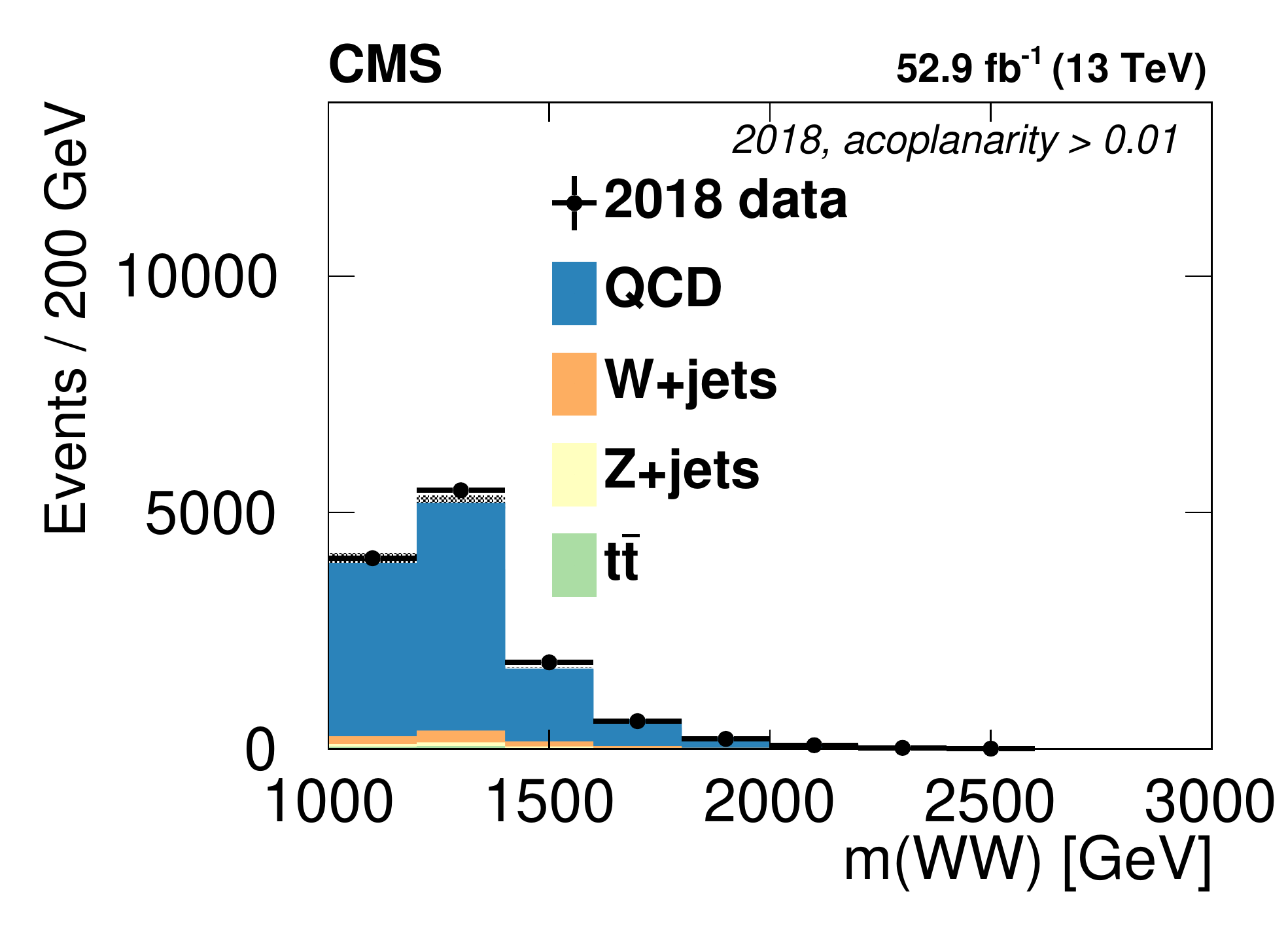}
\includegraphics[width=0.49\textwidth]{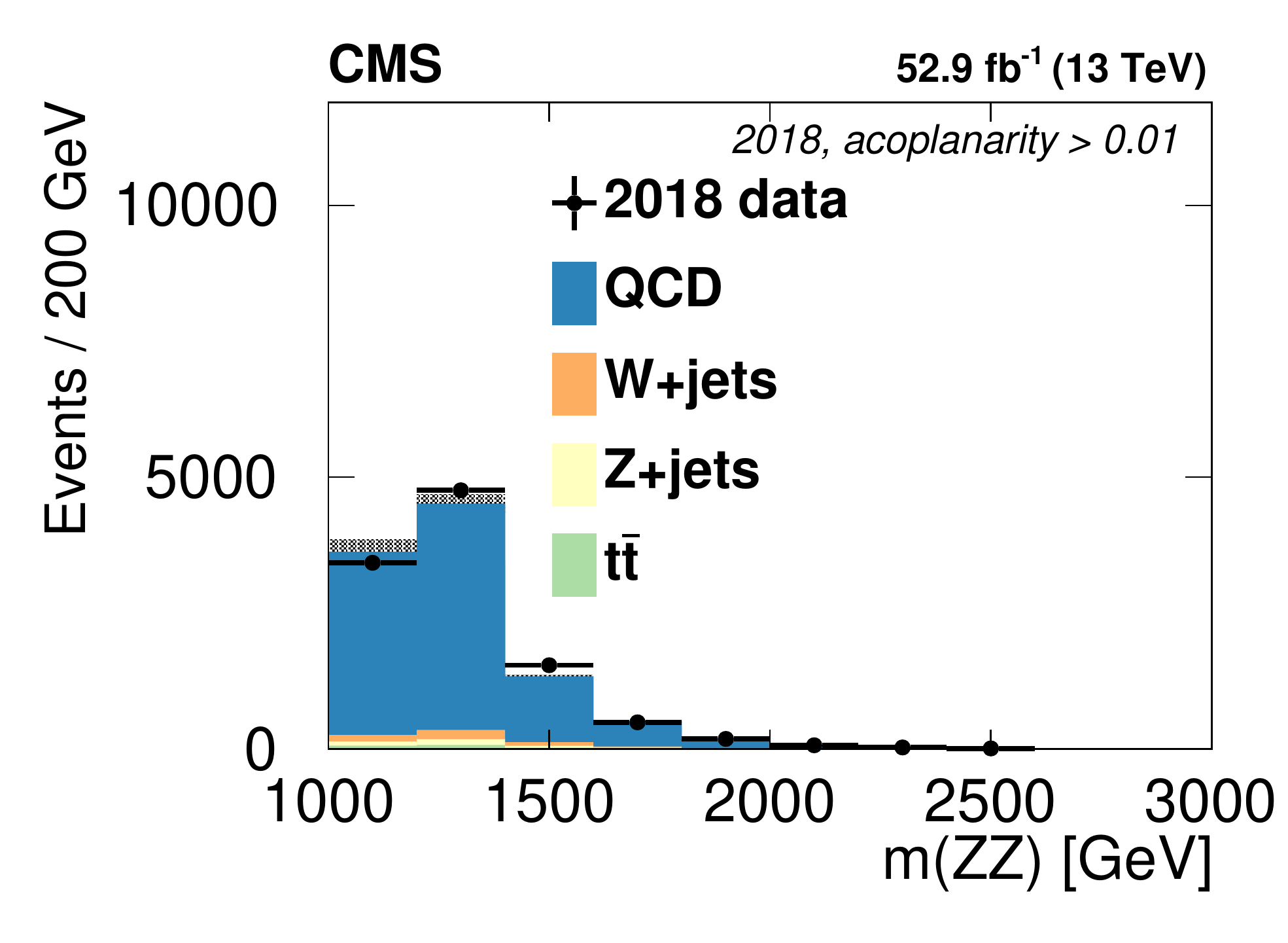}
\caption{Diboson invariant mass distributions in data and simulation in the anti-acoplanarity region ($a>0.01$), with no requirement on the proton matching. The plots from top 
to bottom are for the 2016, 2017, and 2018 data, respectively, with the $\PW\PW$ region in the left column and the $\PZ\PZ$ region in the right column. 
Only statistical uncertainties (dashed grey bands) are shown.}
\label{fig:datamcantiacoplanarity}
\end{figure}

Figure \ref{fig:massrapiditymatchsidebands} illustrates the distribution of data in these regions for the 2018 data with the $\PW\PW$ selection. 
Data selected with the acoplanarity requirement $a > 0.01$ (referred to as the anti-acoplanarity region/method in the following) 
are compared with the predicted background from simulation in Fig.~\ref{fig:datamcantiacoplanarity}, for each of the years and 
in both the $\PW\PW$ and $\PZ\PZ$ mass regions. In general, the data are well reproduced by the simulation, apart from a small excess at low masses in the 
2016 $\PW\PW$ sample. Since the final background estimate is obtained entirely from the data, such minor discrepancies do not impact the results. For 
values of the anomalous couplings near the expected sensitivity of the analysis, the signal contamination in the sideband regions ranges from $< 0.1$ to 
$< 2.7\%$, where we have conservatively used a 95\% CL limit in the various data-taking periods and regions where zero signal MC events actually pass the sideband constraints.

\clearpage

As an alternative background model, we perform the same procedure, except that the selection on the pruned masses of both jets is 
inverted ($m(\Pj) < 60\GeV$ or $m(\Pj) > 107\GeV$), rather than that on the acoplanarity. This method is used only to derive systematic uncertainties.

Finally, as a cross-check we also estimate the background using an event-mixing approach, in which the simulated background events for the central detectors are
mixed with protons from real data, randomly drawn from the jet-triggered data sample. This procedure is repeated 1000 times, and the mean and root-mean-square (RMS) of
the resulting number of events selected in these pseudoexperiments are taken as the background estimate and uncertainty. In the $\delta$ region of the 2016 and 2017 data, 
the number of expected background events is near zero, and the RMS uncertainties can extend to cover negative values. In these cases we instead consider only the mean 
and upper limit on the background prediction, where the upper limit is determined by finding the value that contains 95\% of the pseudoexperiment results. 
This procedure has the drawback of relying on simulation for the central jet samples, and explicitly assumes that the protons always originate from different interactions than the 
jets. Thus, it is not used in the final background estimate, but only as an independent cross-check. Additionally, we use the simulation to perform a closure test of the nominal 
background estimation method, by using the yields in regions B, C, and D to predict the background in region A, and comparing that prediction with the actual number of simulated 
background events in region A. No statistically significant differences are observed.

The resulting background estimates and statistical uncertainties, for all years and signal regions, are shown in Tables~\ref{tab:backgroundestimateswwa} and~\ref{tab:backgroundestimateszza} 
for the default method and the alternative method, based on inversions of the acoplanarity and pruned mass requirements, respectively. In these approaches, the origin of the statistical uncertainties 
is mainly the limited number of events failing the central detector selections but passing the dijet-proton matching (``region B''). The results of the event mixing method 
are also shown, with statistical uncertainties dominated by the limited number of simulated QCD events. When the event mixing method predicts a small number of background events, 
only the mean and upper limit are shown. 

In general, the estimates obtained from the different methods are compatible, given the statistical uncertainties. In all methods the background is dominated by QCD multijet 
production, with other backgrounds contributing ${\leq}8\%$ of the total, according to the Monte Carlo simulation. The background and expected AQGC signal levels in 2018 are much larger than for the other years. In addition to the increase in integrated luminosity, this reflects the improvements in the PPS tracking detectors and ability to reconstruct multiple protons, which significantly improves the signal efficiency but also increases the number of observed background proton tracks. For all years and regions the 
expected SM $\PGg\PGg \to \PW\PW$ signal is less than 0.07 events (Table~\ref{tab:backgroundestimateswwa}). This is primarily due to the high dijet invariant mass requirement 
of $m(\Pj\Pj) > 1126\GeV$ imposed at the preselection level, which reduces the expected signal by more than three orders of magnitude. The less boosted $\PW$ bosons 
from the SM signal are also less likely to be reconstructed as merged jets, which further reduces the acceptance compared with the AQGC signals. In the case of 
$\PGg\PGg \to \PZ\PZ$, the theoretical SM signal is smaller by an additional three orders of magnitude. 

\begin{table}[ht]
\topcaption{Number of background events ($N_\text{evt}$ with statistical uncertainties) expected for all methods in the different $\PW\PW$ analysis regions with 
reconstruction of both signal protons (region $\delta$) or only one signal proton (region $o$). The mean value of the expected number of signal events for one
 anomalous coupling point ($a^{\PW}_{0}/\Lambda^2=5\times 10^{-6}\GeV^{-2}$) and for the SM are also shown for comparison. 
In cases where zero simulated SM events pass the final selection, the value is displayed as a $95\%$ confidence level upper limit.}
\label{tab:backgroundestimateswwa}
\begin{center}
\begin{tabular}{lccccccc}
Number of events & region & $N_\text{evt}$ (2016) & $N_\text{evt}$ (2017) & $N_\text{evt}$ (2018)\\\hline
Anti-acoplanarity sideband   & $\delta$ &  0.4 $\pm$ 0.4 & 1.6 $\pm$ 1.0 & 11.6 $\pm$ 2.6 \\
Anti-pruned mass sideband    & $\delta$ &  0.5 $\pm$ 0.2 & 1.5 $\pm$ 0.3 & 11.3 $\pm$ 0.8 \\
Event mixing                 & $\delta$ &  0.5 (${<}2.2$) & 1.8 (${<}4.2$) & 14.3 $\pm$ 8.9 \\[\cmsTabSkip]
Expected signal              & $\delta$ &  1.7           & 2.2           & 16.1 \\
($a^{\PW}_{0}/\Lambda^2=5\times 10^{-6}\GeV^{-2}$) & & & & \\
Expected signal (SM)              & $\delta$ & 0.006          & $< 0.05$        & 0.03 \\[\cmsTabSkip]
Anti-acoplanarity sideband   & $o$ &  1.4 $\pm$ 0.9 & 10.0 $\pm$ 3.2 & 41.4 $\pm$ 5.7 \\
Anti-pruned mass sideband    & $o$ &  2.5 $\pm$ 0.8 & 7.1 $\pm$ 1.3 & 43.0 $\pm$ 3.0 \\
Event mixing                 & $o$ &  2.4 $\pm$ 1.9 & 8.4 $\pm$ 6.3 & 49 $\pm$ 13 \\[\cmsTabSkip]
Expected signal              & $o$ &  1.5           & 1.7           & 16.8  \\
($a^{\PW}_{0}/\Lambda^2=5\times 10^{-6}\GeV^{-2}$) & & & & \\                                                                                         
Expected signal (SM)         & $o$ &  0.005         & ${<}0.05$           & ${< }0.07$ \\[\cmsTabSkip]
\end{tabular}
\end{center}
\end{table}

\begin{table}
\topcaption{
Number of background events ($N_\text{evt}$ with statistical uncertainties) expected for all methods in the different $\PZ\PZ$ analysis regions with 
reconstruction of both signal protons (region $\delta$) or only one signal proton (region $o$). The mean value of the expected number of the expected signal for one 
anomalous coupling point ($a^{\PZ}_{0}/\Lambda^2=1\times 10^{-5}\GeV^{-2}$) is also shown for comparison.} 
\label{tab:backgroundestimateszza}
\begin{center}
\begin{tabular}{lccccccc}
Number of events & region & $N_\text{evt}$ (2016) & $N_\text{evt}$ (2017) & $N_\text{evt}$ (2018)\\\hline
Anti-acoplanarity sideband    & $\delta$ &  1.5 $\pm$ 1.1 & 1.6 $\pm$ 0.8 & 14.2 $\pm$ 3.0 \\
Anti-pruned mass sideband     & $\delta$ &  0.4 $\pm$ 0.2 & 0.9 $\pm$ 0.2 &  9.9 $\pm$ 0.9 \\
Event mixing                  & $\delta$ &  0.5 ($< 2.1$) & 1.5 ($< 3.6$) & 11.6 $\pm$ 9.4 \\[\cmsTabSkip]
Expected signal               & $\delta$ &  1.3           & 1.4           & 9.0 \\
($a^{\PZ}_{0}/\Lambda^2=1\times 10^{-5}\GeV^{-2}$) &      &               &     &     \\[\cmsTabSkip]

Anti-acoplanarity sideband    & $o$ &  1.5 $\pm$ 1.1 & 3.7 $\pm$ 1.5 & 37.4 $\pm$ 5.6 \\
Anti-pruned mass sideband     & $o$ &  2.1 $\pm$ 0.8 & 5.4 $\pm$ 1.3 & 41.7 $\pm$ 3.1 \\
Event mixing                  & $o$ &  2.0 $\pm$ 1.8 & 6.3 $\pm$ 5.1 & 42 $\pm$ 16 \\[\cmsTabSkip]
Expected signal               & $o$ &  1.0           & 1.6           & 12.8 \\
($a^{\PZ}_{0}/\Lambda^2=1\times 10^{-5}\GeV^{-2}$) & &               &     &     \\
\end{tabular}
\end{center}
\end{table}

\section{Systematic uncertainties}
We estimate systematic uncertainties in the signal prediction due to the jet energy scale, proton $\xi$ measurement, proton reconstruction efficiency, and
integrated luminosity. The effect of the jet energy scale is evaluated by shifting the energy of both jets in the event up or down by the uncertainty, and
recomputing the expected signal yields. The resulting uncertainties in the signal yield depend on the data-taking period and sample, but typically range from a few 
percent up to $10\%$. Possible discrepancies between data and MC regarding the pruned mass and the $\tau_{21}$ variable are found to be negligible (below 1\%) using a tag and probe method.

The systematic uncertainties related to the proton $\xi$ reconstruction in the two arms are assumed to be uncorrelated. There are two reasons for 
this assumption: (i) the RP alignment relative to the beam, and the uncertainty in the LHC magnet strengths that determine the optics, are independent for the two colliding beams;
and (ii) a significant part of the uncertainty is related to the stability over time. In the case of the alignment, 
this is a fill-by-fill effect, that fluctuates in both directions around the mean whenever the Roman pots are moved in or out. The optics uncertainty includes 
a component related to the LHC crossing angle, which was adjusted several times per fill (in 2017) or continuously throughout the fill (in 2018). 
An approximation is made that the expected signal is less than 1 event per fill (which is the expected case for the present analysis),
 and therefore these components of the uncertainty are not correlated event-by-event.

The effect of the proton reconstruction uncertainty is evaluated by shifting the value of $\xi$ for each simulated signal proton by an amount 
drawn from a Gaussian distribution with width equal to the uncertainty~\cite{propaper}, and recomputing the expected signal yield. Because of the tight matching requirements between protons and jets, this is one of the largest systematic uncertainties in the analysis, with values typically of the order of $30\%$ of the expected signal yield. 
The proton reconstruction efficiency uncertainties are based on comparing results obtained with different control samples and efficiency estimation procedures~\cite{propaper}. The total efficiency uncertainty per arm is $10\%$ in 2016; in 2017 and 2018 the improved detectors and methods led to uncertainties between 2 and $3\%$ per arm \cite{propaper}. The efficiency uncertainties for the two arms are assumed uncorrelated, and are summed in quadrature to obtain the event uncertainty. The integrated luminosity uncertainties are estimated as $1.2\%$~\cite{lum1}, $2.3\%$~\cite{lum2}, and $2.5\%$~\cite{lum3} for the 2016, 2017, and 2018 samples, respectively. The uncertainty includes both a part that is uncorrelated between years, and parts that 
are correlated between 2016 and 2017, and between 2016, 2017, and 2018. The overall uncertainty for the 2016--2018 period for data collected with PPS is 1.8\%.  

The theoretical uncertainty in the signal prediction includes contributions from the simulation sample size, and from the proton survival probability,
\ie, the probability that the colliding protons do not break up due to soft interactions between their spectator partons. The 
statistical uncertainty in the simulation varies with the assumed anomalous coupling value and year simulated. For the points closest to the expected sensitivity, which 
determine the numerical value of the limits, it is typically in the range of 2--10\%. For the smallest coupling values simulated, which are far from the expected sensitivity, it is 
typically 25\% in the 2016 samples where the integrated luminosity of the data is low. The uncertainty in the proton survival probability is assumed to be 10\%~\cite{sur}. 

The systematic uncertainty in the background has two contributions. The first is the statistical uncertainty in the normalization, based on the nominal acoplanarity sideband method.
The second is the dependence of the background estimate on the choice of the sideband region. This is estimated as the full difference between the central values of the acoplanarity 
sideband method, and the pruned mass sideband method. The first uncertainty, which depends mainly on the amount of sideband data, ranges from ${\approx}15-20\%$ in the 2018 data, to ${>}100\%$ in the 2016 data. 
The second uncertainty also has a significant statistical component, which ranges from a few percent in the 2018 data, to 80\% in the 2016 data.

\section{Signal extraction and results}

The high-mass exclusive $\PGg\PGg \to \PW\PW$ and $\PGg\PGg \to \PZ\PZ$ signals are extracted using a binned maximum-likelihood fit in a total of twelve bins: one for each of the three years of data taking, times two for the $\PW\PW$ and $\PZ\PZ$ regions, times two for the events with either reconstruction of both signal protons (region $\delta$) or only one signal proton (region $o$). In most cases, systematic uncertainties are included as log-normal nuisance parameters~\cite{Conway:2011in}. 
When the estimate of the systematic uncertainty is based on a sample with less than 10 events, Poisson nuisances are used instead; this applies to some of the background statistical uncertainties derived from sideband regions, and some of the simulation statistical 
errors for very small couplings. The signal is estimated for each of the anomalous couplings, with
all other couplings fixed to zero.

\begin{figure}[h!t]
\centering
\includegraphics[width=0.8\textwidth]{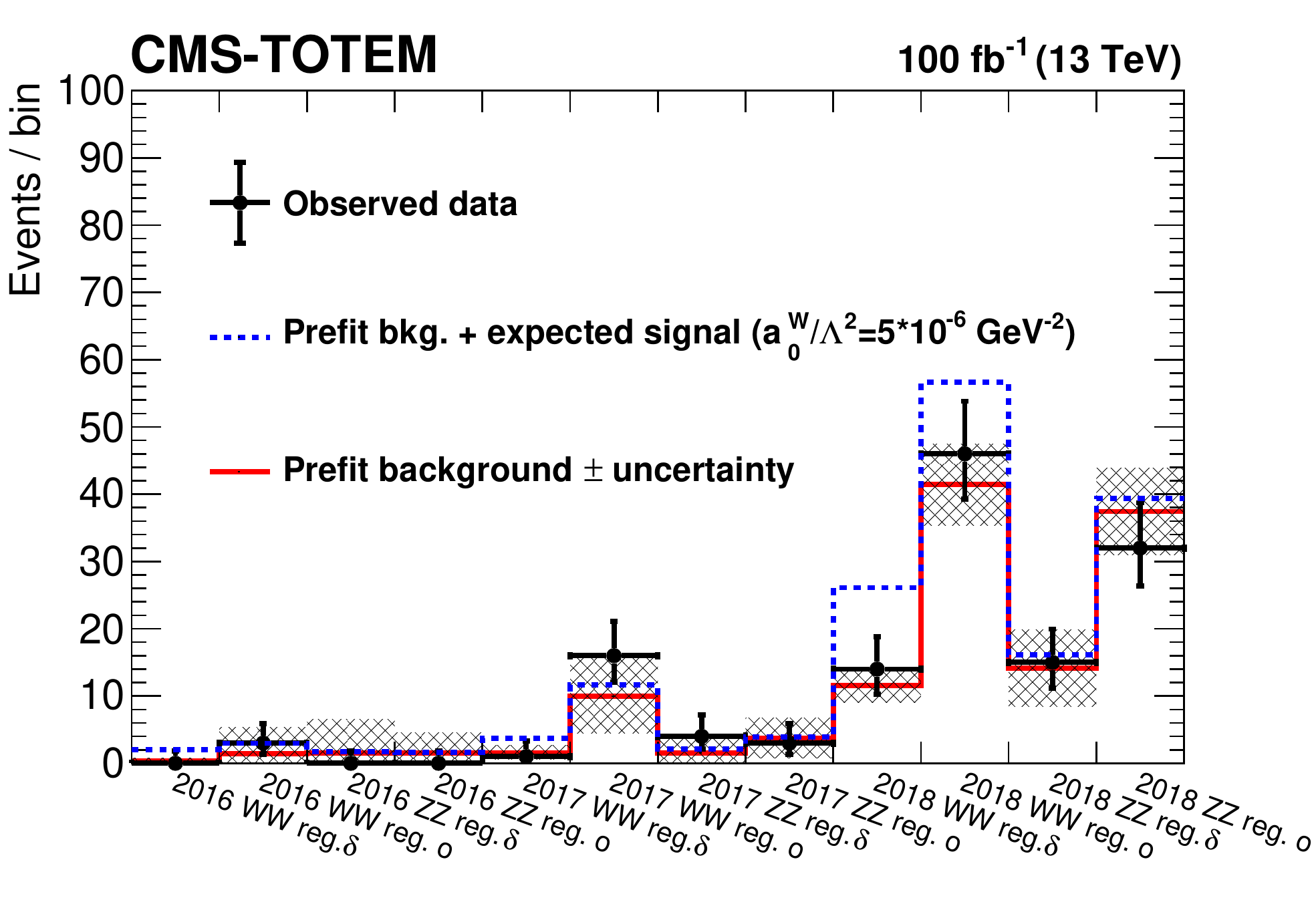}
\includegraphics[width=0.8\textwidth]{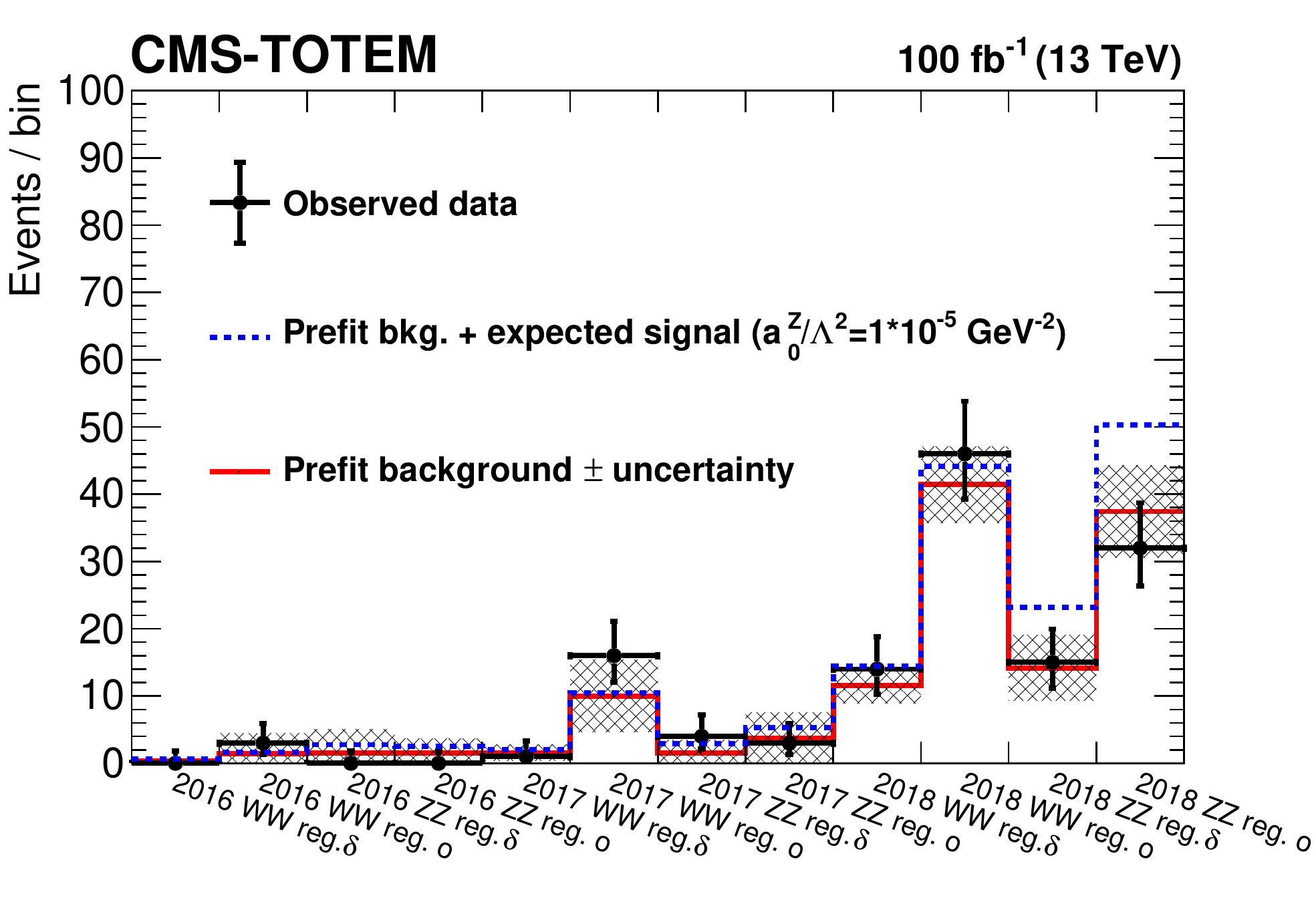}
\caption{Observed data and expected number of background events in each signal region. Hypothetical AQGC signals are also shown. The histogram with solid lines indicates
the number expected for only background, with uncertainties shown by the shaded band. The dashed-line histogram shows the number for background plus assumed signals 
with $a^{\PW}_{0}/\Lambda^2=5\times 10^{-6}\GeV^{-2}$ (upper) or $a^{\PZ}_{0}/\Lambda^2=1\times 10^{-5}\GeV^{-2}$ (lower). The histograms and 
uncertainties are shown prior to the binned maximum-likelihood fit described in the text. The shaded band indicates the uncertainty in the background estimate, 
while the vertical bars on the points represent the statistical uncertainty in the observed data.}
\label{fig:unrollsignalregions}
\end{figure}

Figure~\ref{fig:unrollsignalregions} shows the number of observed events compared with the expected background and a hypothetical signal, in each bin of the 
analysis. The backgrounds and observed data are compared with a signal having nonzero $\PW\PW$ anomalous couplings, slightly above the expected sensitivity of the 
analysis. Small (${\leq}1 \sigma$) excesses are seen in ``region $o$'' of the $\PW\PW$ for all years, while small deficits are seen in the $\PZ\PZ$ channel.
None of these excesses or deficits in the data is significant.

\subsection{AQGC limits}

\begin{figure}[ht]
\centering
\includegraphics[width=0.49\linewidth]{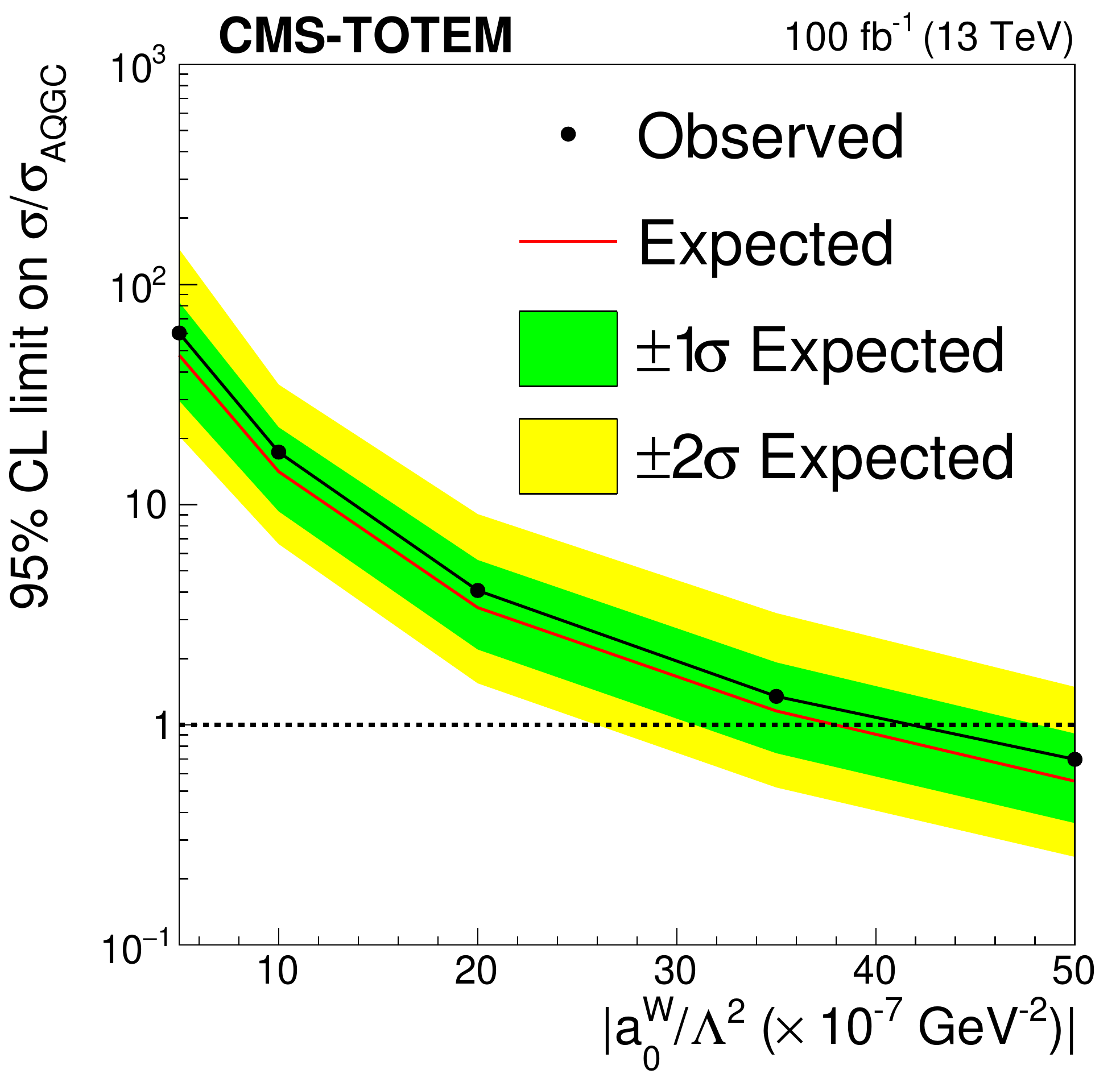}
\includegraphics[width=0.49\linewidth]{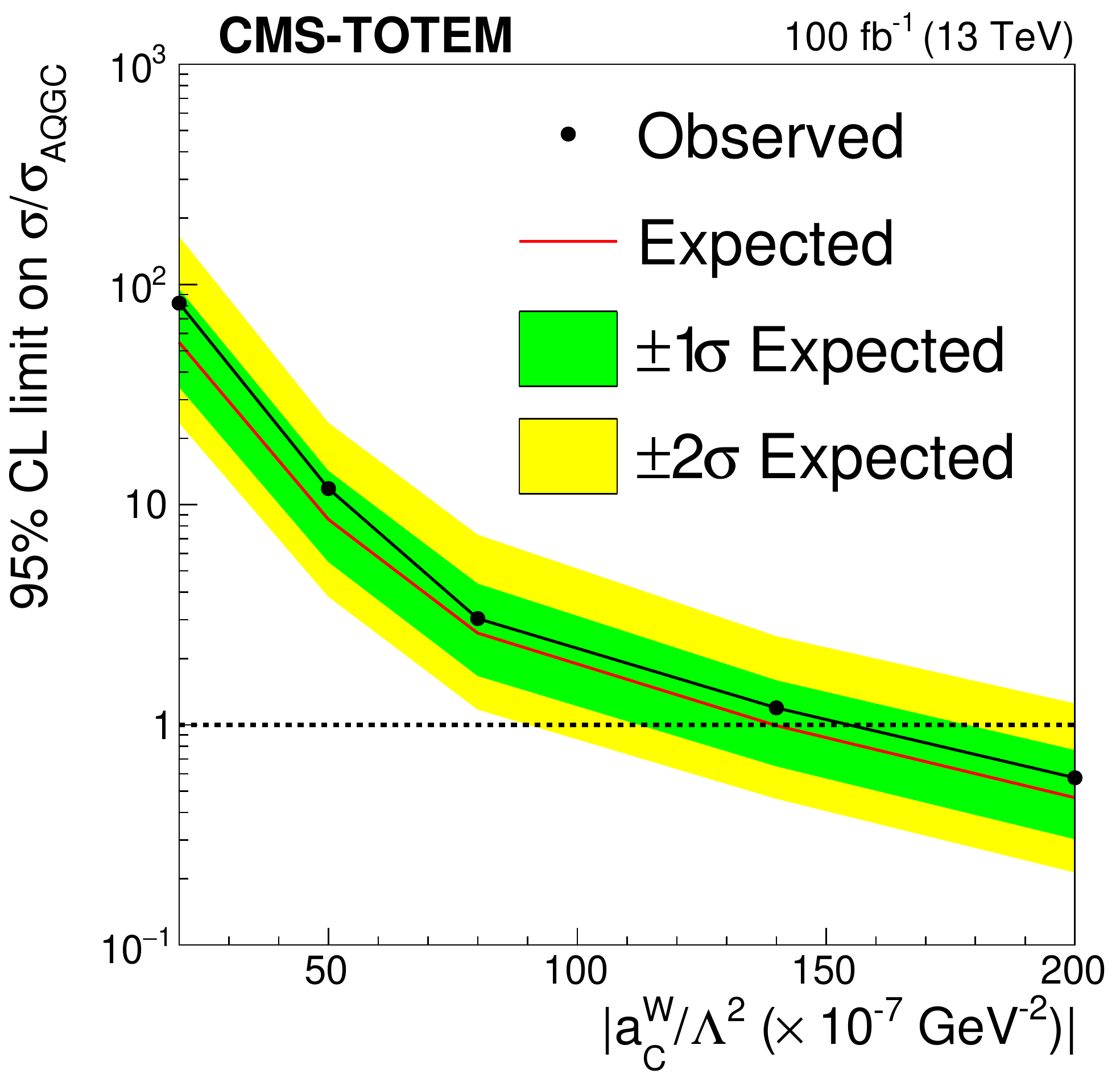}
\includegraphics[width=0.49\linewidth]{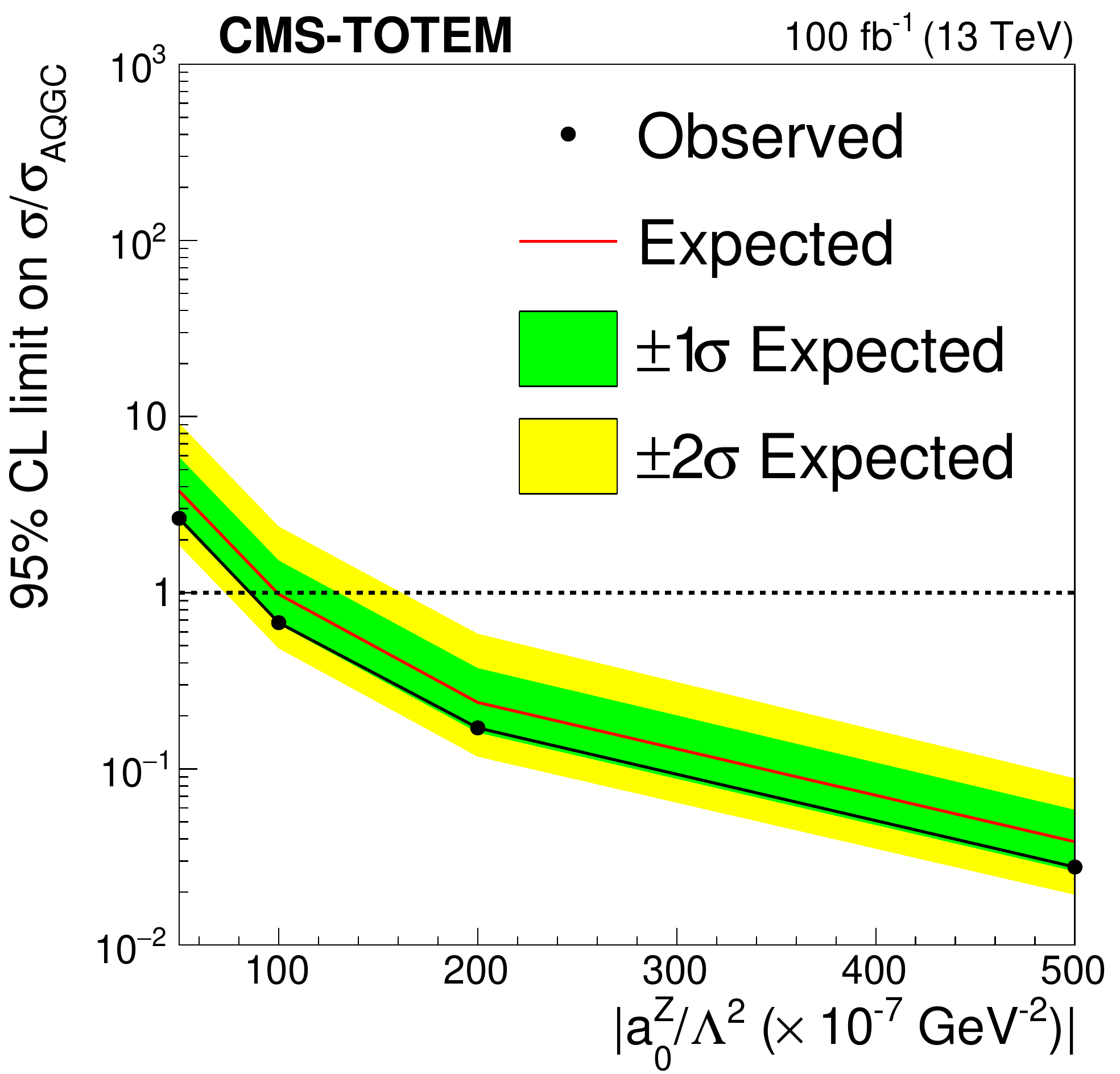}
\includegraphics[width=0.49\linewidth]{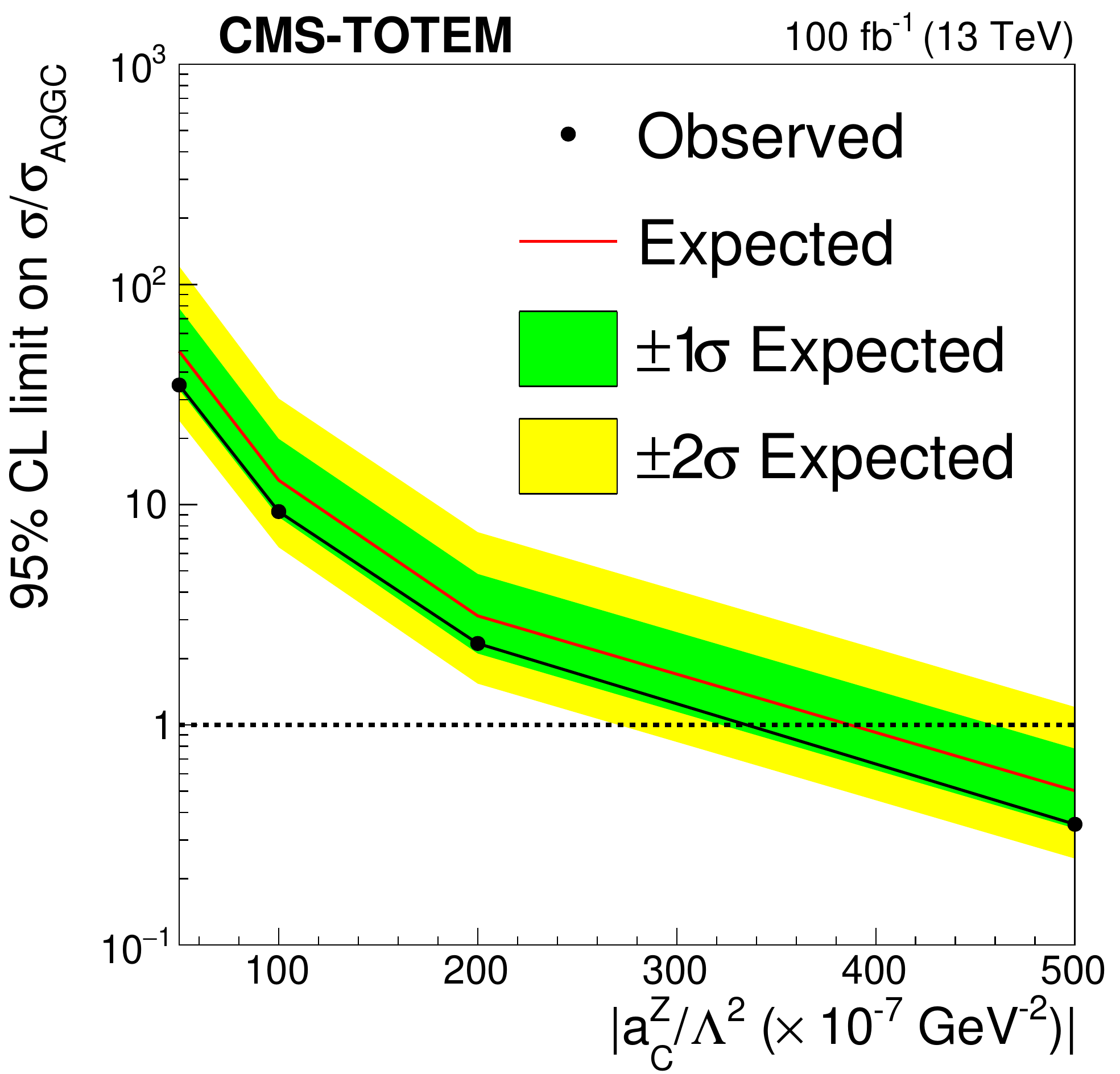}
\caption{Expected and observed upper limits on the AQGC operators $a^{\PW}_{0}/\Lambda^{2}$ (upper left), $a^{\PW}_{C}/\Lambda^{2}$ (upper right), 
$a^{\PZ}_{0}/\Lambda^{2}$ (lower left), $a^{\PZ}_{C}/\Lambda^{2}$ (lower right), with no unitarization. The $y$ axis shows the limit on the ratio 
of the observed cross section to the cross section predicted for each anomalous coupling value ($\sigma_\mathrm{AQGC}$). The graphs are not fully representative close to the values of the 95
limit due to technical plotting features; the 95\% CL upper limits quoted in Table~\ref{tab:aqgclimits} correspond to
the exact values.} 
\label{fig:limits}
\end{figure}

The resulting expected and observed $95\%$ confidence level (\CL) upper limits on the AQGC operators are shown in Fig.~\ref{fig:limits}. 

For large values of anomalous couplings, the predicted cross section becomes unphysically large at high masses of the produced diboson system, and violates
partial wave unitarity. We estimate the sensitivity of the limits to this effect by calculating the energy of $\PGg\PGg$ 
collisions at which unitarity is violated for the expected limits~\cite{Eboli:2003nq}, and then apply a ``clipping'' procedure~\cite{Kalinowski:2018oxd} to remove 
the simulated signal above that value. The background estimate and the data are left unchanged in this procedure. In Ref.~\cite{CMS:2020gfh} 
the SM contribution to the signal was retained above the clipping threshold; in the current analysis the total SM signal is negligible, and is not 
included. We then rederive the limits with the clipping applied.

In the $\PW\PW$ channel, unitarity violation occurs at a diboson mass of $\Lambda \approx1.4\TeV$. By clipping the signal model at that value, we 
obtain new expected limits that are approximately 30 to 40\% higher than the unitarity-violating limits. In the $\PZ\PZ$ channel, where the expected limits 
are higher, unitarity violation occurs at approximately 1.1\TeV for both $a^{\PZ}_{0}/\Lambda^{2}$ and $a^{\PZ}_{C}/\Lambda^{2}$. In this case, because of the invariant mass
threshold imposed by the jet triggers, there is no value of clipping for which unitarity is preserved. The full list of $95\%$ \CL upper limits with and 
without clipping is shown in Table~\ref{tab:aqgclimits}.

\begin{table}
\topcaption{Limits on LEP-like dimension-6 anomalous quartic gauge coupling parameters~\cite{OPAL:2004voc}, with and without unitarization via a clipping~\cite{CMS:2020gfh} procedure.}
\label{tab:aqgclimits}
\begin{center}
\renewcommand{\arraystretch}{1.3}
\begin{tabular}{ccc}
Coupling & Observed (expected)  & Observed (expected)\\[-5pt]
& 95\% \CL upper limit  & 95\% \CL upper limit \\[-5pt]
         & No clipping                              & Clipping at 1.4\TeV \\\hline
$\abs{a^{\PW}_{0}/\Lambda^{2}}$ & $ 4.3 ~(3.9) \times 10^{-6}\GeV^{-2}$ & $ 5.2 ~(5.1) \times 10^{-6}\GeV^{-2}$ \\
$\abs{a^{\PW}_{C}/\Lambda^{2}}$ & $ 1.6 ~(1.4) \times 10^{-5}\GeV^{-2}$ & $ 2.0 ~(2.0) \times 10^{-5}\GeV^{-2}$ \\ 
$\abs{a^{\PZ}_{0}/\Lambda^{2}}$ & $ 0.9 ~(1.0) \times 10^{-5}\GeV^{-2}$ & \NA \\
$\abs{a^{\PZ}_{C}/\Lambda^{2}}$ & $ 4.0 ~(4.5) \times 10^{-5}\GeV^{-2}$ & \NA \\
\end{tabular}
\end{center}
\end{table}

\begin{figure}[ht]
\centering
\includegraphics[width=0.49\textwidth]{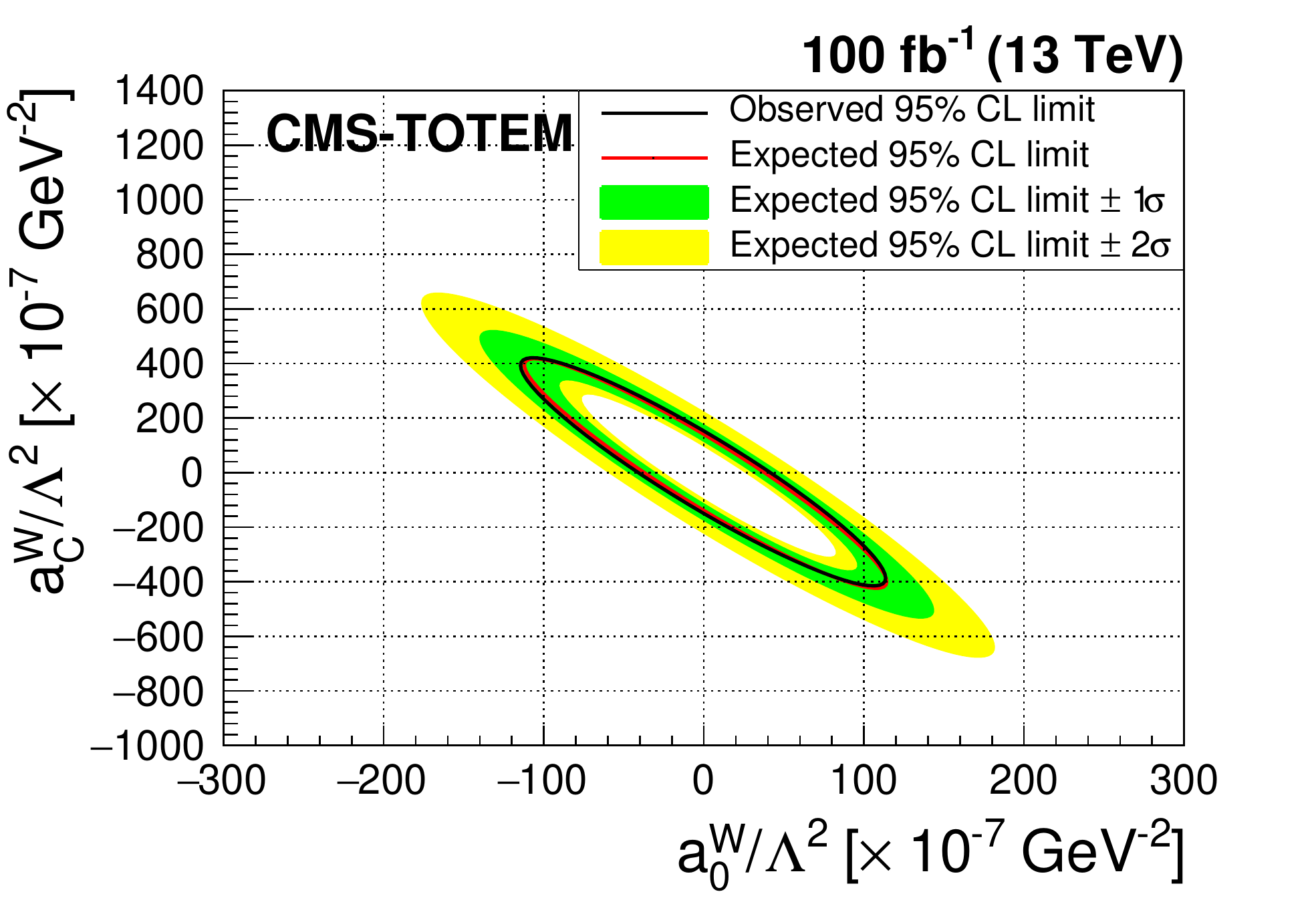}
\includegraphics[width=0.49\textwidth]{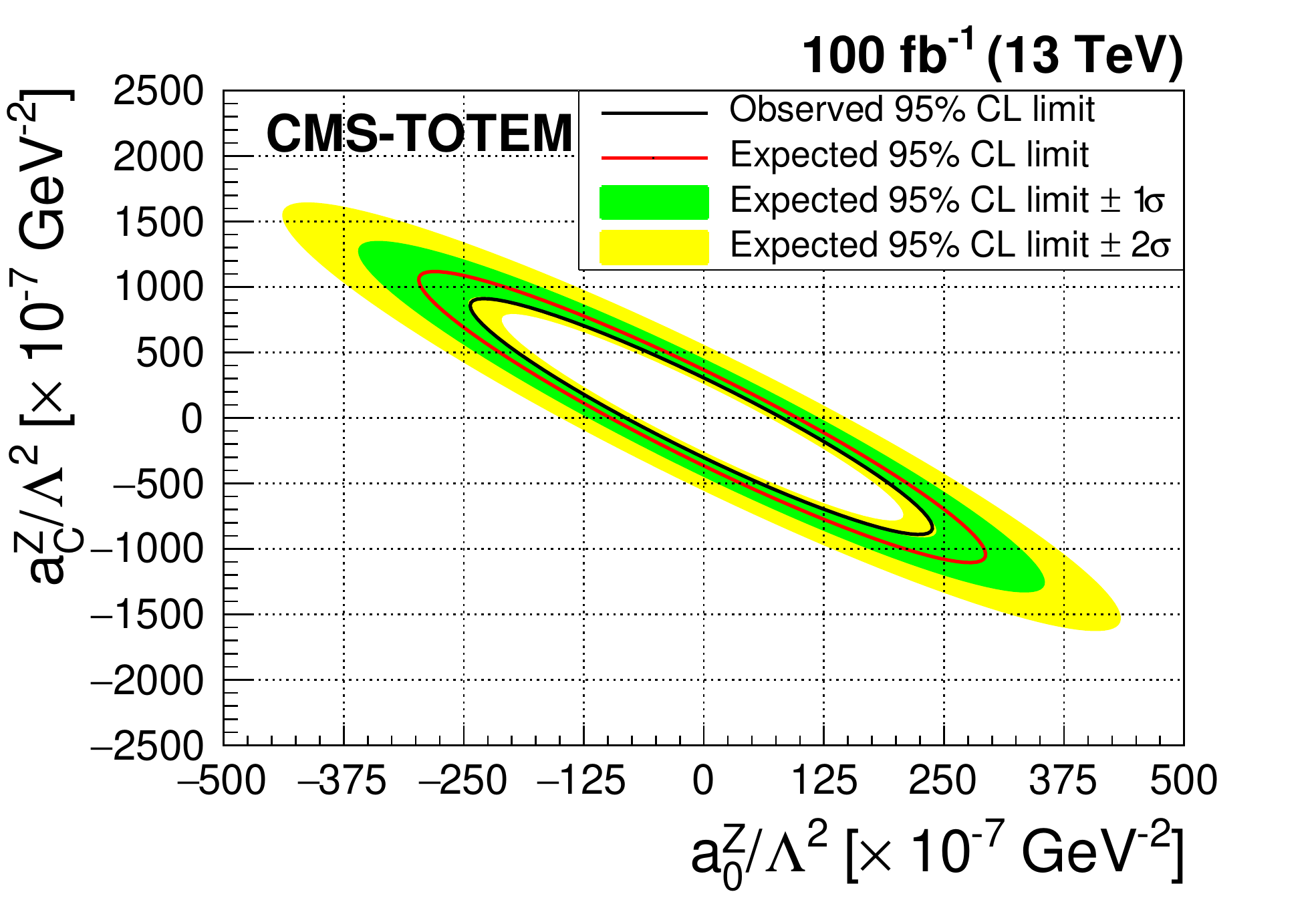}

\includegraphics[width=0.49\textwidth]{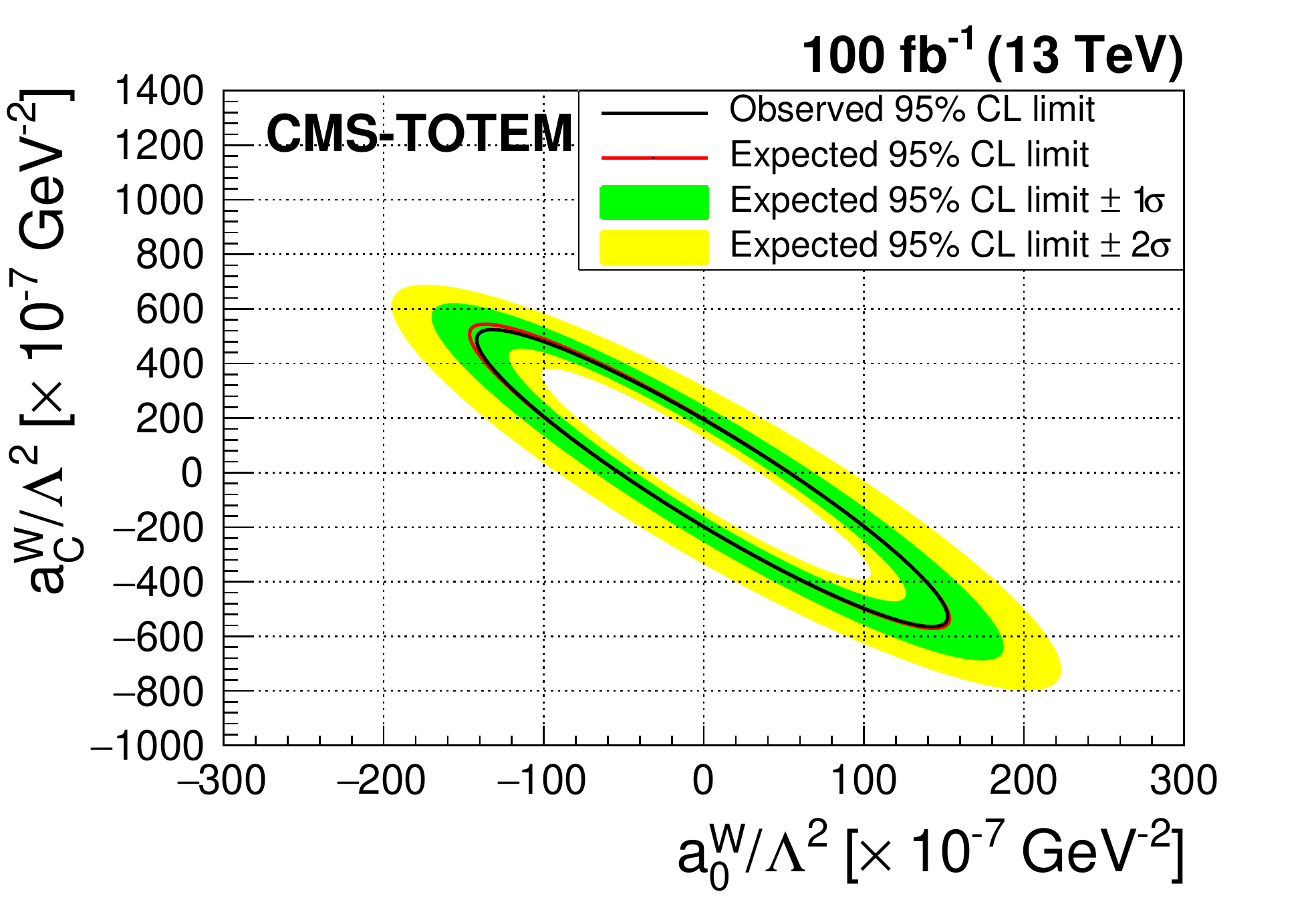}
\caption{Expected and observed limits in the two-dimensional plane of $a^{\PW}_{0}/\Lambda^{2}$ vs $a^{\PW}_{C}/\Lambda^{2}$ (upper left), $a^{\PZ}_{0}/\Lambda^{2}$ vs. $a^{\PZ}_{C}/\Lambda^{2}$ (upper right), 
and $a^{\PW}_{0}/\Lambda^{2}$ vs. $a^{\PW}_{C}/\Lambda^{2}$ with unitarization imposed by clipping the signal model at 1.4\TeV (lower).}
\label{fig:twodlimits}
\end{figure}

We also obtain limits in the two-dimensional planes of $a^{\PW}_{0}/\Lambda^{2}$ vs. $a^{\PW}_{C}/\Lambda^{2}$ and $a^{\PZ}_{0}/\Lambda^{2}$ vs. $a^{\PZ}_{C}/\Lambda^{2}$ by 
fitting the limits on the ratio of observed to predicted cross sections to an analytical model for the dependence of the cross section on the AQGC values. The results are 
shown in Fig.~\ref{fig:twodlimits}.
The $a^{\PW}_{0}/\Lambda^{2}$ vs. $a^{\PW}_{C}/\Lambda^{2}$ limits are shown with and without clipping the signal model at 1.4\TeV. 
Compared with the limits on the $\abs{a^{\PW}_{0}/\Lambda^{2}}$ and $\abs{a^{\PW}_{C}/\Lambda^{2}}$ couplings obtained with LHC Run 1 data~\cite{ATLAS:2016lse,CMS:2016rtz}, 
where unitarization was imposed via a dipole form factor, the unitarized limits obtained here represent an improvement by a factor ${\approx}15$--20. 

\clearpage

\subsection{Translation to linear dimension-8 AQGCs}

Many recent anomalous coupling searches quote limits on dimension-8 linear operators alone. These operators are classified in several categories, depending on 
whether they contain only covariant derivatives of the Higgs field, only field strength tensors, or both (``mixed'' operators, with 
couplings denoted $f_{M}$)~\cite{Eboli:2016kko}. In the case of processes involving photons, the $a^{\PW,\PZ}_{0,C}$ operators can be translated into a linear combination 
of ``mixed'' dimension-8 $f_{M,i}\,(i = \text{0--7})$ operators~\cite{Eboli:2016kko}. For the $a^{\PW}_{0}$ coupling, the relationship reads~\cite{Eboli:2016kko}:

\begin{linenomath}
\begin{equation}
a^{\PW}_{0} = - \frac{m_{\PW}}{\pi \alpha_{\text{em}}}\left[s^2_w \frac{f_{M,0}}{\Lambda^2} + 2 c^2_w \frac{f_{M,2}}{\Lambda^2} + s_w c_w \frac{f_{M,4}}{\Lambda^2}\right].
\end{equation}
\end{linenomath}

Here $m_{\PW}$ is the \PW boson mass, $\alpha_{\text{em}}$ is the fine structure constant, and $s_w$ and $c_w$ represent the sine and cosine of the weak mixing angle, respectively. 
By further assuming that anomalous contributions to $\PW\PW \PZ\PGg$ vanish, an additional constraint of $f_{M,0} + 2f_{M,2}$ is obtained~\cite{Belanger:1999aw,CMS:2014cdf}, 
allowing $a^{\PW}_{0}$ to be written in terms of only $f_{M,0}$ and $f_{M,4}$. In order to compare with other results, we scan the values of $f_{M,0}$ and $f_{M,4}$ for which 
the limit on $a^{\PW}_{0}$ is satisfied. We find the results shown in Table~\ref{tab:d8limits}, evaluated at the point where the other coupling is zero.

\begin{table}[ht]
\topcaption{Conversion of limits on $a^{\PW}_0$ to dimension-8 $f_{M,i}$ operators, using the assumption of vanishing $\PW\PW\PZ\PGg$ couplings to eliminate some parameters. When quoting limits on one of the operators, the other is fixed to zero. The results for $\abs{f_{M,0}/\Lambda^{4}}$ and $\abs{f_{M,4}/\Lambda^{4}}$ are shown with and without clipping of the signal model at 1.4\TeV, when the other parameter is fixed to the SM value of zero.}
\label{tab:d8limits}
\begin{center}
\renewcommand{\arraystretch}{1.3}

\begin{tabular}{ccc}
Coupling & Observed (expected)   & Observed (expected) \\[-5pt]
&95\% \CL upper limit & 95\% \CL upper limit\\[-5pt]
         & No clipping                              & Clipping at 1.4\TeV \\\hline
$\abs{f_{M,0}/\Lambda^{4}}$  & $ 16.2 ~(14.7)\TeV^{-4}$ & $ 19.5 ~(19.2)\TeV^{-4}$ \\ 
$\abs{f_{M,4}/\Lambda^{4}}$  & $ 90.9 ~(82.6)\TeV^{-4}$ & $ 110 ~(108)\TeV^{-4}$ \\
\end{tabular}
\end{center}
\end{table}

The results are compared with other vector boson scattering studies at 13\TeV that are sensitive to the same
operators~\cite{CMS:2021gme,CMS:2021gme,CMS:2021gme}. When the conversion is performed assuming vanishing $\PW\PW\PZ\PGg$ couplings, the
unitarity-violating limits on $\abs{f_{M,0}/\Lambda^{4}}$ and $\abs{f_{M,4}/\Lambda^{4}}$ are several times looser
than those quoted in other measurements. After the clipping, the results for $\abs{f_{M,0}/\Lambda^{4}}$ are similar to the best limits obtained from
vector boson scattering in the same sign $\PWpm\PWpm$ and $\PW\PZ$ final states in CMS~\cite{CMS:2020gfh}. The milder effect of the unitarization
in this analysis is due to the upper mass limit of ${\approx}2\TeV$ for reconstructing the protons, imposed by the LHC collimator apertures.
This already suppresses most of the unphysical high-mass part of the signal model, before applying any unitarization procedure.

\begin{table}[ht]
\topcaption{Conversion of limits on $a^{\PW}_0$ and $a^{\PW}_C$ to dimension-8 $f_{M,i}$ operators, using the assumption that all $f_{M,i}$ except one are equal to zero.
The results are shown with and without clipping of the signal model at 1.4\TeV.}
\label{tab:d8limitsotherszero}
\begin{center}
\renewcommand{\arraystretch}{1.3}
\begin{tabular}{ccc}
Coupling & Observed (expected)   & Observed (expected) \\[-5pt]
& 95\% \CL upper limit & 95\% \CL upper limit\\[-5pt]
         & No clipping                              & Clipping at 1.4\TeV \\\hline
$\abs{f_{M,0}/\Lambda^{4}}$ & $ 66.0 ~(60.0)\TeV^{-4}$ & $ 79.8 ~(78.2)\TeV^{-4}$ \\
$\abs{f_{M,1}/\Lambda^{4}}$ & $ 245.5 ~(214.8)\TeV^{-4}$ & $ 306.8 ~(306.8)\TeV^{-4}$ \\
$\abs{f_{M,2}/\Lambda^{4}}$ & $ 9.8 ~(9.0)\TeV^{-4}$ & $ 11.9 ~(11.8)\TeV^{-4}$ \\
$\abs{f_{M,3}/\Lambda^{4}}$ & $ 73.0 ~(64.6)\TeV^{-4}$ & $ 91.3 ~(92.3)\TeV^{-4}$ \\
$\abs{f_{M,4}/\Lambda^{4}}$ & $ 36.0 ~(32.9)\TeV^{-4}$ & $ 43.5 ~(42.9)\TeV^{-4}$ \\
$\abs{f_{M,5}/\Lambda^{4}}$ & $ 67.0 ~(58.9)\TeV^{-4}$ & $ 83.7 ~(84.1)\TeV^{-4}$ \\
$\abs{f_{M,7}/\Lambda^{4}}$ & $ 490.9 ~(429.6)\TeV^{-4}$ & $ 613.7 ~(613.7)\TeV^{-4}$ \\
\end{tabular}
\end{center}
\end{table}

Alternatively, all dimension-8 $f_{M,i}$ except one are set equal to zero. In this case, the results of the conversion are shown in Table~\ref{tab:d8limitsotherszero}.
When the conversion is performed by setting all other couplings to zero, the limits on $\abs{f_{M,0}/\Lambda^{4}}$ are significantly looser, the limits on 
$\abs{f_{M,4}/\Lambda^{4}}$ are somewhat more restrictive, and tight constraints are obtained on $\abs{f_{M,2}/\Lambda^{4}}$.

\subsection{Fiducial cross sections}

In addition to the limits on different anomalous coupling parameters, we derive upper limits on the cross section for an AQGC-like signal in 
the $\Pp\Pp\to\Pp\PW\PW\Pp$ and $\Pp\Pp\to\Pp\PZ\PZ\Pp$ channels. The limits are obtained for a fiducial region of $0.04 < \xi < 0.20$ and diboson invariant mass $m(\PV\PV) > 1\TeV$, and 
correspond to the diboson production cross section before decays into hadrons. The signal simulation is used to extrapolate from the measurement region 
to the fiducial region. As with the AQGC limits, the cross section limits are obtained for each channel 
separately, assuming zero signal in the other channel. After verifying that the signal efficiency and acceptance do not depend strongly on the exact 
value of the AQGCs, for $0.04 < \xi < 0.20$ and $m(\PV\PV) > 1000\GeV$ we measure the following upper exclusion limits at 95\% \CL: $\sigma(\Pp\Pp \to \Pp \PW\PW \Pp) < 67 (53^{+34}_{-19})\unit{fb}$ and $\sigma(\Pp\Pp \to \Pp \PZ\PZ \Pp) < 43 (62^{+33}_{-20})\unit{fb}$, where the expected limit and its one standard deviation uncertainty are shown in parentheses.
Since the simulation is used to extrapolate the results from the measurement region to the fiducial region, the cross section limits given are valid for signals with an AQGC-like $\xi$ and mass dependence. The limits cannot therefore be applied to models with a very different mass and $\xi$ behaviour. 
\section{Summary}

A first search for exclusive high-mass $\PGg\PGg \to \PW\PW$ and $\PGg\PGg \to \PZ\PZ$ production with reconstructed forward protons has 
been performed, in final states with hadronically decaying $\PW$ or $\PZ$ bosons, using 100\fbinv of data collected in 13\TeV proton-proton collisions. No significant excess is found over the standard model background prediction. The resulting limits are interpreted in terms of nonlinear dimension-6 and linear dimension-8 anomalous 
quartic gauge couplings (AQGC). 

The unitarized dimension-6 $\PGg\PGg \PW\PW$ AQGC limits are approximately 15--20 times more stringent than those obtained from the $\PGg\PGg \to \PW\PW$ process without 
proton detection using the LHC Run 1 data~\cite{CMS:2016rtz,ATLAS:2016lse}. In the analyses without proton detection~\cite{CMS:2016rtz,ATLAS:2016lse,ATLAS:2020iwi}, 
larger SM signals were measured at lower invariant $\PW\PW$ masses. However, since enhancements due to AQGCs come mainly from the high energy region of 
$\gamma\gamma$ interactions, the analysis described here is more sensitive than the limits obtained in those publications. 
In addition, the use of proton detectors results in a theoretically cleaner result, by reducing the dependency on the unitarization scheme, and reducing the contribution of signal events where the protons dissociate.
The derived dimension-8 limits are close to those obtained from same-sign $\PW\PW$ and $\PW\PZ$ scattering at 13\TeV after unitarization, for the case when the $\PW\PW\PZ\PGg$ coupling is suppressed. The limits on $\PGg\PGg \PZ\PZ$ anomalous couplings are the first 
obtained from the exclusive $\PGg\PGg \to \PZ\PZ$ channel. New limits are placed on the cross section in a fiducial region of $0.04 < \xi < 0.20$ and diboson invariant mass $m(\PV\PV) > 1\TeV$
of $\PGg\PGg \to \PW\PW$ and $\PGg\PGg \to \PZ\PZ$  production with intact forward protons.

Tabulated results are provided in HEPData~\cite{HEPData}.

\begin{acknowledgments}
We congratulate our colleagues in the CERN accelerator departments for the excellent performance of the LHC and thank the technical and administrative staffs at CERN and at other CMS and TOTEM institutes for their contributions to the success of the common CMS-TOTEM effort. In addition, we gratefully acknowledge the computing centers and personnel of the Worldwide LHC Computing Grid and other centers for delivering so effectively the computing infrastructure essential to our analyses. Finally, we acknowledge the enduring support for the construction and operation of the LHC, the CMS and TOTEM detectors, and the supporting computing infrastructure provided by the following funding agencies: BMBWF and FWF (Austria); FNRS and FWO (Belgium); CNPq, CAPES, FAPERJ, FAPERGS, and FAPESP (Brazil); MES and BNSF (Bulgaria); CERN; CAS, MoST, and NSFC (China); MINCIENCIAS (Colombia); MSES and CSF (Croatia); RIF (Cyprus); SENESCYT (Ecuador); MoER, ERC PUT and ERDF (Estonia); Academy of Finland, Magnus Ehrnrooth Foundation, MEC, HIP, and Waldemar von Frenckell Foundation (Finland); CEA and CNRS/IN2P3 (France); BMBF, DFG, and HGF (Germany); GSRI (Greece); the Circles of Knowledge Club and NKFIH (Hungary); DAE and DST (India); IPM (Iran); SFI (Ireland); INFN (Italy); MSIP and NRF (Republic of Korea); MES (Latvia); LAS (Lithuania); MOE and UM (Malaysia); BUAP, CINVESTAV, CONACYT, LNS, SEP, and UASLP-FAI (Mexico); MOS (Montenegro); MBIE (New Zealand); PAEC (Pakistan); MES and NSC (Poland); FCT (Portugal); MESTD (Serbia); MCIN/AEI and PCTI (Spain); MOSTR (Sri Lanka); Swiss Funding Agencies (Switzerland); MST (Taipei); MHESI and NSTDA (Thailand); TUBITAK and TENMAK (Turkey); NASU (Ukraine); STFC (United Kingdom); DOE and NSF (USA).
    
\hyphenation{Rachada-pisek} Individuals have received support from the Marie-Curie program and the European Research Council and Horizon 2020 Grant, contract Nos.\ 675440, 724704, 752730, 758316, 765710, 824093, 884104, and COST Action CA16108 (European Union); the Leventis Foundation; the Alfred P.\ Sloan Foundation; the Alexander von Humboldt Foundation; the Belgian Federal Science Policy Office; the Fonds pour la Formation \`a la Recherche dans l'Industrie et dans l'Agriculture (FRIA-Belgium); the Agentschap voor Innovatie door Wetenschap en Technologie (IWT-Belgium); the F.R.S.-FNRS and FWO (Belgium) under the ``Excellence of Science -- EOS" -- be.h project n.\ 30820817; the Beijing Municipal Science \& Technology Commission, No. Z191100007219010; the Ministry of Education, Youth and Sports (MEYS) of the Czech Republic; Svenska Kulturfonden (Finland); the Hellenic Foundation for Research and Innovation (HFRI), Project Number 2288 (Greece); the Deutsche Forschungsgemeinschaft (DFG), under Germany's Excellence Strategy -- EXC 2121 ``Quantum Universe" -- 390833306, and under project number 400140256 - GRK2497; the Hungarian Academy of Sciences, the New National Excellence Program - \'UNKP, the NKFIH research grants K 124845, K 124850, K 128713, K 128786, K 129058, K 131991, K 133046, K 138136, K 143460, K 143477, 2020-2.2.1-ED-2021-00181, and TKP2021-NKTA-64 (Hungary); the Council of Science and Industrial Research, India; the Latvian Council of Science; the Ministry of Education and Science, project no. 2022/WK/14, and the National Science Center, contracts Opus 2021/41/B/ST2/01369 and 2021/43/B/ST2/01552 (Poland); the Funda\c{c}\~ao para a Ci\^encia e a Tecnologia, grant CEECIND/01334/2018 (Portugal); MCIN/AEI/10.13039/501100011033, ERDF ``a way of making Europe", and the Programa Estatal de Fomento de la Investigaci{\'o}n Cient{\'i}fica y T{\'e}cnica de Excelencia Mar\'{\i}a de Maeztu, grant MDM-2017-0765 and Programa Severo Ochoa del Principado de Asturias (Spain); the Chulalongkorn Academic into Its 2nd Century Project Advancement Project, and the National Science, Research and Innovation Fund via the Program Management Unit for Human Resources \& Institutional Development, Research and Innovation, grant B05F650021 (Thailand); the Kavli Foundation; the Nvidia Corporation; the SuperMicro Corporation; the Welch Foundation, contract C-1845; and the Weston Havens Foundation (USA).  
\end{acknowledgments}

\bibliography{auto_generated}
\cleardoublepage \appendix\section{The CMS Collaboration \label{app:collab}}\begin{sloppypar}\hyphenpenalty=5000\widowpenalty=500\clubpenalty=5000
\cmsinstitute{Yerevan Physics Institute, Yerevan, Armenia}
{\tolerance=6000
A.~Tumasyan\cmsAuthorMark{1}\cmsorcid{0009-0000-0684-6742}
\par}
\cmsinstitute{Institut f\"{u}r Hochenergiephysik, Vienna, Austria}
{\tolerance=6000
W.~Adam\cmsorcid{0000-0001-9099-4341}, J.W.~Andrejkovic, T.~Bergauer\cmsorcid{0000-0002-5786-0293}, S.~Chatterjee\cmsorcid{0000-0003-2660-0349}, K.~Damanakis\cmsorcid{0000-0001-5389-2872}, M.~Dragicevic\cmsorcid{0000-0003-1967-6783}, A.~Escalante~Del~Valle\cmsorcid{0000-0002-9702-6359}, P.S.~Hussain\cmsorcid{0000-0002-4825-5278}, M.~Jeitler\cmsAuthorMark{2}\cmsorcid{0000-0002-5141-9560}, N.~Krammer\cmsorcid{0000-0002-0548-0985}, L.~Lechner\cmsorcid{0000-0002-3065-1141}, D.~Liko\cmsorcid{0000-0002-3380-473X}, I.~Mikulec\cmsorcid{0000-0003-0385-2746}, P.~Paulitsch, F.M.~Pitters, J.~Schieck\cmsAuthorMark{2}\cmsorcid{0000-0002-1058-8093}, R.~Sch\"{o}fbeck\cmsorcid{0000-0002-2332-8784}, D.~Schwarz\cmsorcid{0000-0002-3821-7331}, S.~Templ\cmsorcid{0000-0003-3137-5692}, W.~Waltenberger\cmsorcid{0000-0002-6215-7228}, C.-E.~Wulz\cmsAuthorMark{2}\cmsorcid{0000-0001-9226-5812}
% ...
\par}
\vskip\cmsinstskip
\dag:~Deceased\\
$^{1}$Also at Yerevan State University, Yerevan, Armenia\\
$^{2}$Also at TU Wien, Vienna, Austria\\
% ...

\section{The TOTEM Collaboration\label{app:totem}}
\newcommand{\AddAuthor}[2]{#1$^{#2}$,\ }
\newcommand\AddAuthorLast[2]{#1$^{#2}$}
\noindent
\AddAuthor{G.~Antchev\cmsorcid{0000-0003-3210-5037}}{a}
\AddAuthor{P.~Aspell}{8}
\AddAuthor{I.~Atanassov\cmsorcid{0000-0002-5728-9103}}{a}
\AddAuthor{V.~Avati}{7,8}
\AddAuthor{J.~Baechler}{8}
\AddAuthor{C.~Baldenegro~Barrera\cmsorcid{0000-0002-6033-8885}}{10}
\AddAuthor{V.~Berardi\cmsorcid{0000-0002-8387-4568}}{4a,4b}
\AddAuthor{M.~Berretti\cmsorcid{0000-0003-4122-8282}}{2a}
\AddAuthor{V.~Borshch\cmsorcid{0000-0002-5479-1982}}{11}
\AddAuthor{E.~Bossini\cmsorcid{0000-0002-2303-2588}}{6a}
\AddAuthor{U.~Bottigli\cmsorcid{0000-0002-0666-3433}}{6c}
\AddAuthor{M.~Bozzo\cmsorcid{0000-0002-1715-0457}}{5a,5b}
\AddAuthor{H.~Burkhardt}{8}
\AddAuthor{F.S.~Cafagna\cmsorcid{0000-0002-7450-4784}}{4a}
\AddAuthor{M.G.~Catanesi\cmsorcid{0000-0002-2987-7688}}{4a}
\AddAuthor{M.~Csan\'{a}d\cmsorcid{0000-0002-3154-6925}}{3a,b}
\AddAuthor{T.~Cs\"{o}rg\H{o}\cmsorcid{0000-0002-9110-9663}}{3a,3b}
\AddAuthor{M.~Deile\cmsorcid{0000-0001-5085-7270}}{8}
\AddAuthor{F.~De~Leonardis}{4a,4c}
\AddAuthor{M.~Doubek\cmsorcid{0000-0002-0026-9558}}{1c}
\AddAuthor{D.~Druzhkin}{8,11}
\AddAuthor{K.~Eggert}{9}
\AddAuthor{V.~Eremin}{e}
\AddAuthor{A.~Fiergolski}{8}
\AddAuthor{F.~Garcia\cmsorcid{0000-0002-4023-7964}}{2a}
\AddAuthor{V.~Georgiev}{1a}
\AddAuthor{S.~Giani}{8}
\AddAuthor{L.~Grzanka\cmsorcid{0000-0002-3599-854X}}{7}
\AddAuthor{J.~Hammerbauer}{1a}
\AddAuthor{T.~Isidori\cmsorcid{0000-0002-7934-4038}}{10}
\AddAuthor{V.~Ivanchenko\cmsorcid{0000-0002-1844-5433}}{11}
\AddAuthor{M.~Janda\cmsorcid{0000-0001-5736-6183}}{1c}
\AddAuthor{A.~Karev}{8}
\AddAuthor{J.~Ka\v{s}par\cmsorcid{0000-0001-5639-2267}}{1b,8}
\AddAuthor{B.~Kaynak\cmsorcid{0000-0003-3857-2496}}{c}
\AddAuthor{J.~Kopal\cmsorcid{0000-0001-9751-7409}}{8}
\AddAuthor{V.~Kundr\'{a}t\cmsorcid{0000-0003-2868-5550}}{1b}
\AddAuthor{S.~Lami\cmsorcid{0000-0001-9492-0147}}{6a}
\AddAuthor{R.~Linhart}{1a}
\AddAuthor{C.~Lindsey}{10}
\AddAuthor{M.V.~Lokaj\'{i}\v{c}ek\cmsorcid{0000-0002-2052-1220}$^{\textrm{\dag,}}$}{1b}
\AddAuthor{L.~Losurdo\cmsorcid{0000-0002-4964-7951}}{6c}
\AddAuthor{F.~Lucas~Rodr\'{i}guez}{8}
\AddAuthor{M.~Macr\'{i}}{\dag,5a}
\AddAuthor{M.~Malawski\cmsorcid{0000-0001-6005-0243}}{7}
\AddAuthor{N.~Minafra\cmsorcid{0000-0003-4002-1888}}{10}
\AddAuthor{S.~Minutoli}{5a}
\AddAuthor{K.~Misan\cmsorcid{0009-0007-2416-042X}}{7}
\AddAuthor{T.~Naaranoja\cmsorcid{0000-0001-5797-7929}}{2a,2b}
\AddAuthor{F.~Nemes\cmsorcid{0000-0002-1451-6484}}{3a,3b,8}
\AddAuthor{H.~Niewiadomski}{9}
\AddAuthor{T.~Nov\'{a}k\cmsorcid{0000-0001-6253-4356}}{3b}
\AddAuthor{E.~Oliveri\cmsorcid{0000-0002-0832-6975}}{8}
\AddAuthor{F.~Oljemark\cmsorcid{0000-0003-0121-2761}}{2a,2b}
\AddAuthor{M.~Oriunno}{d}
\AddAuthor{K.~\"{O}sterberg\cmsorcid{0000-0003-4807-0414}}{2a,2b}
\AddAuthor{P.~Palazzi\cmsorcid{0000-0002-4861-391X}}{8}
\AddAuthor{V.~Passaro\cmsorcid{0000-0003-0802-4464}}{4a,4c}
\AddAuthor{Z.~Peroutka}{1a}
\AddAuthor{J.~Proch\'{a}zka\cmsorcid{0000-0002-2774-2245}}{1b}
\AddAuthor{M.~Quinto\cmsorcid{0000-0002-6363-6132}}{4a,4b}
\AddAuthor{E.~Radermacher}{8}
\AddAuthor{E.~Radicioni\cmsorcid{0000-0002-2231-1067}}{4a}
\AddAuthor{F.~Ravotti\cmsorcid{0000-0002-5709-6934}}{8}
\AddAuthor{C.~Royon\cmsorcid{0000-0002-7672-9709}}{10}
\AddAuthor{G.~Ruggiero}{8}
\AddAuthor{H.~Saarikko}{2a,2b}
\AddAuthor{V.D.~Samoylenko}{e}
\AddAuthor{A.~Scribano\cmsorcid{0000-0002-4338-6332}}{6a,10}
\AddAuthor{J.~\v{S}irok\'{y}}{1a}
\AddAuthor{J.~Smajek}{8}
\AddAuthor{W.~Snoeys\cmsorcid{0000-0003-3541-9066}}{8}
\AddAuthor{R.~Stefanovitch}{8}
\AddAuthor{J.~Sziklai}{3a}
\AddAuthor{C.~Taylor\cmsorcid{0000-0001-6816-8051}}{9}
\AddAuthor{E.~Tcherniaev\cmsorcid{0000-0002-3685-0635}}{11}
\AddAuthor{N.~Turini\cmsorcid{0000-0002-9395-5230}}{6c}
\AddAuthor{O.~Urban}{1a}
\AddAuthor{V.~Vacek\cmsorcid{0000-0001-9584-0392}}{1c}
\AddAuthor{O.~Vavroch}{1a}
\AddAuthor{J.~Welti}{2a,2b}
\AddAuthor{J.~Williams\cmsorcid{0000-0002-9810-7097}}{10}
\AddAuthorLast{J.~Zich}{1a}

\vskip 4pt plus 4pt
\let\thefootnote\relax
\newcommand{\AddInstitute}[2]{${}^{#1}$#2\\}
\newcommand{\AddExternalInstitute}[2]{\footnote{${}^{#1}$ #2}}
\noindent
\dag Deceased\\
\AddInstitute{1a}{University of West Bohemia, Pilsen, Czech Republic}
\AddInstitute{1b}{Institute of Physics of the Academy of Sciences of the Czech Republic, Prague, Czech Republic}
\AddInstitute{1c}{Czech Technical University, Prague, Czech Republic}
\AddInstitute{2a}{Helsinki Institute of Physics, University of Helsinki, Helsinki, Finland}
\AddInstitute{2b}{Department of Physics, University of Helsinki, Helsinki, Finland}
\AddInstitute{3a}{Wigner Research Centre for Physics, RMKI, Budapest, Hungary}
\AddInstitute{3b}{MATE Institute of Technology KRC, Gy\"{o}ngy\"{o}s, Hungary}
\AddInstitute{4a}{INFN Sezione di Bari, Bari, Italy}
\AddInstitute{4b}{Dipartimento Interateneo di Fisica di Bari, University of Bari, Bari, Italy}
\AddInstitute{4c}{Dipartimento di Ingegneria Elettrica e dell'Informazione --- Politecnico di Bari, Bari, Italy}
\AddInstitute{5a}{INFN Sezione di Genova, Genova, Italy}
\AddInstitute{5b}{Universit\`{a} degli Studi di Genova, Genova, Italy}
\AddInstitute{6a}{INFN Sezione di Pisa, Pisa, Italy}
\AddInstitute{6c}{Universit\`{a} degli Studi di Siena and Gruppo Collegato INFN di Siena, Siena, Italy}
\AddInstitute{7}{Akademia G\'{o}rniczo-Hutnicza (AGH) University of Science and Technology, Krakow, Poland}
\AddInstitute{8}{CERN, Geneva, Switzerland}
\AddInstitute{9}{Case Western Reserve University, Department of Physics, Cleveland, Ohio, USA}
\AddInstitute{10}{The University of Kansas, Lawrence, Kansas, USA}
\AddInstitute{11}{Authors affiliated with an institute or an international laboratory covered by a cooperation agreement with CERN}
\AddExternalInstitute{a}{INRNE-BAS, Institute for Nuclear Research and Nuclear Energy, Bulgarian Academy of Sciences, Sofia, Bulgaria}
\AddExternalInstitute{b}{Department of Atomic Physics, E\"{o}tv\"{o}s Lor\'{a}nd University, Budapest, Hungary}
\AddExternalInstitute{c}{Istanbul University, Istanbul, Turkey}
\AddExternalInstitute{d}{SLAC, Stanford University, California, USA}
\AddExternalInstitute{e}{Authors affiliated with an institute or an international laboratory covered by a cooperation agreement with CERN}\end{sloppypar}
\end{document}